\documentclass[aps,prx,reprint,twocolumn,superscriptaddress,longbibliography]{revtex4-2}
\usepackage{amsmath}
\usepackage{amsfonts}
\usepackage{MnSymbol}
\usepackage{xcolor}
\usepackage{braket}
\usepackage{graphicx}
\usepackage{siunitx}
\usepackage{hyperref}
\hypersetup{
	colorlinks=true,
	linkcolor=blue,
	filecolor=blue,
	citecolor = blue,      
	urlcolor=cyan,
}
\newcommand{\figref}[1]{Fig.~\ref{#1}}

\newcommand{\secref}[1]{Sec.~\ref{#1}}
\newcommand{\eqeqref}[1]{Eq.~(\ref{#1})}

\newcommand{\balpha}{\boldsymbol{\alpha}}

\begin{document}

\title{Tuning between collective dynamics and scrambling through initial-state symmetry alignment
in long-range spin models}

\author{Sridevi Kuriyattil}
\email{sridevi.kuriyattil@physics.ox.ac.uk}
\affiliation{Clarendon Laboratory, University of Oxford, Parks Road, Oxford OX1 3PU, United Kingdom}
\author{Johannes Kombe}
\affiliation{Department of Physics, SUPA and University of Strathclyde, Glasgow G4 0NG, United Kingdom}
\author{Andrew J. Daley}
\affiliation{Clarendon Laboratory, University of Oxford, Parks Road, Oxford OX1 3PU, United Kingdom}
\author{Pablo M. Poggi}
\email{pablo.poggi@strath.ac.uk}
\affiliation{Department of Physics, SUPA and University of Strathclyde, Glasgow G4 0NG, United Kingdom}

\begin{abstract}
Long-range interactions in many-body quantum systems give rise to a wide range of dynamical phenomena, from collective dynamics well described by integrable mean-field models with symmetric all-to-all couplings, to fast scrambling of quantum information from disordered long-range interactions. Here, we investigate scenarios in which both regimes emerge within a single deterministic, translation-invariant XY spin model  with long-range couplings. We consider a sparse powers-of-two (PWR2) graph, where each spin interacts with a logarithmic number of neighbors, and a power-law–decaying model. By altering the initial state via local perturbations on a completely symmetric state, we show that the dynamics changes from collective to scrambling as witnessed by bipartite and multipartite entanglement measures. We characterize this collective-to-scrambling crossover using Lyapunov exponents in the semiclassical regime, and show that the spectral properties of the coupling graph adjacency matrix provide a very efficient tool to predict the onset of fast scrambling. Our results offer a unified framework for controlling the switch between collective and scrambling dynamics in a setup accessible in experiments with neutral atoms in cavities or in tweezer arrays.  
\end{abstract}
\maketitle
\section{Introduction}

Quantum many-body dynamics is strongly shaped by the range and structure of interactions. Long-range couplings, in particular, allow correlations and information to spread across a system in ways that can differ fundamentally from their short-range counterparts \cite{LiebRobinson1972,pappalardiScramblingEntanglementSpreading2018,DefenuLerosePappalardi2023,Richerme2014,tran2020}. One particularly striking phenomenon enabled by such interactions is fast scrambling, a process in which initially localised quantum information spreads throughout a quantum system at a maximal rate, becoming inaccessible to local measurements \cite{hosurChaosQuantumChannels2016,lashkari2013towards,page1993average,sekino2008fast}. In such systems, correlations become highly nonlocal in a time scale $t_* \propto \log(N)$, and the resulting states are strongly entangled, generically exploring the full $2^N$ dimensional Hilbert space of an $N$ qubit system. \cite{yao2016interferometric,Greg_tree_2019,marino2019cavity,bentsenFastScramblingSparse2019,li2020fast,belyansky2020minimal,hashizumeDeterministicFastScrambling2021}. The exponential growth of this space with system size typically makes such states increasingly difficult to represent and simulate (with some exceptions, see \cite{Sachdev1993-SYK,Maldacena_2016,KimDiracFS2021,kitaevSimpleModelQuantum2015,claeys2021}).

However, not all long-range interacting many body systems explore an exponentially large state space. In particular, when the Hamiltonian is invariant under the exchange of any two particles, the evolution of initially symmetric states is restricted to a permutationally symmetric manifold, whose dimension grows linearly with system size $N$. For systems of spin-1/2 particles, the dynamics can be represented in terms of a collective degree of freedom and the resulting states, while possessing restricted bipartite entanglement, tend to be good resource states for quantum-enhanced sensing \cite{Kitagawa_1993,ma_quantum_2011,Gross2012,PezzeSmerziOberthalerSchmiedTreutlein2018}. The prototypical example of such a Hamiltonian is the One Axis Twisting (OAT) model that features infinite range Ising interactions between the spins \cite{Kitagawa_1993,ma_quantum_2011,munoz2023}. Recent efforts have focused on understanding to what extent spin systems with finite range interactions can mimic the OAT dynamics, for instance in Rydberg atom arrays featuring dipole-dipole interactions \cite{Bornet_2023}, among others \cite{eckner2023, hines2023}.

\begin{figure}
    \centering
    \includegraphics[width=1.0\linewidth]{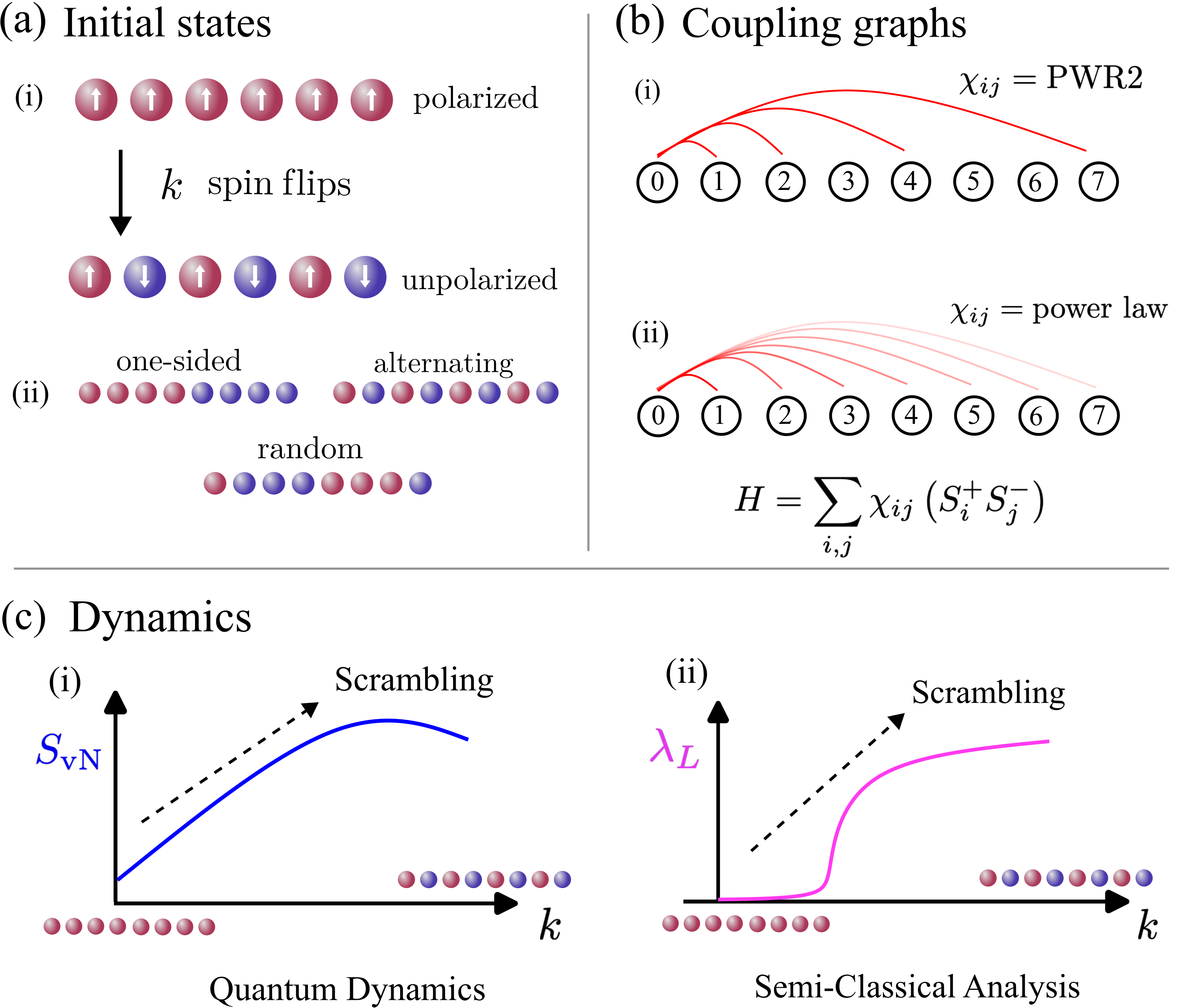}
    \caption{(a) Initial states obtained by flipping spins from a fully polarized configuration. Unflipped and flipped spins are shown as red and blue spheres, respectively. We consider three flipping protocols: one-sided, alternating, and random. (b) The initial states are evolved under the XY Hamiltonian, for two different one-dimensional coupling graphs with adjacency matrix $\chi$. In graph (i)
    spins separated by distances that are powers of two are connected; in graph (ii), the interaction strength decays with distance as $|i-j|^{\alpha}$. (c) Quantum and semiclassical dynamics are characterized by the long-time bipartite von Neumann entanglement entropy (i) and the Lyapunov exponent (ii), respectively. In both cases, the corresponding quantities increase with the number of spin flips $k$, signalling a crossover between collective and scrambling dynamics.}
    \label{fig:overview_fig}
\end{figure}

The existence of such distinct dynamical behaviors in long-range interacting systems raises the fundamental question of how these regimes are connected, and whether they can emerge as different dynamical manifestations of the same underlying Hamiltonian.  
In this work we identify realizable systems in which it is possible to switch from a collective to a scrambling regime merely by local control of the initial state.
We demonstrate this crossover in two distinct one-dimensional Hamiltonians featuring long-range spin-exchange interactions: the powers-of-two (PWR2) model \cite{gubserContinuumLimitsSparse2018,Greg_tree_2019,MIPT_tomo_2022,hashizumeTunableGeometriesSparse2022,Sridevi_Onset_2023,SK_metrology_2025,gunning2025geometrydriventransitionssparselongrange} and the power-law (PL) model \cite{Dyson1969PhaseTransition,DuttaBhattacharjee2001LongRangeIsingRotor,DefenuLerosePappalardi2023,Pupillo_Kitaev_PL,Schachenmayer2013_EG}. 
Starting from a fully spin-polarized, permutationally symmetric state, we consider states with $k$ flipped spins, as illustrated in Fig.~\ref{fig:overview_fig} (a), and study the resulting dynamics. As $k$ is increased, the system departs from collective behavior and progressively develops signatures of scrambling in both its quantum and semiclassical dynamics.  
Our findings show that--well beyond trivial regimes where the initial state and the Hamiltonian share a particular symmetry--the dynamical regime is not determined by the Hamiltonian alone. Instead, both collective and scrambling dynamics can emerge within the same interacting model, tuned through the initial state's degree of permutational symmetry breaking. 
We note also that the combination of long-range interactions and local state engineering considered here is accessible in a range of experimental platforms, including atoms in optical cavities \cite{Periwal_2021,qin_2019,Hung_2016,vaidya_2018,Mivehvar_2021}, tweezer arrays \cite{Bluvstein_2022,Bluvstein_2023,Graham2022,Xu2024,Evered2023,Bernien2017}, and trapped ions \cite{Islam2013,Britton_2012,Richerme_2014,Guo2024,Feng2023,Smith2016}.

We further show that key aspects of this dynamical crossover can be understood efficiently in terms of the underlying coupling graph. In the semiclassical regime, we show that the spectral properties of the graph adjacency matrix directly determine the sensitivity of different configurations to small perturbations. In the quantum regime, we introduce an initial-state-dependent graph Laplacian and relate its minimum spectral gap to the onset of the collective-to-scrambling transition, extending previous studies \cite{SK_metrology_2025,Perlin_SS_2020}. Together, these results provide a framework for understanding and optimising the visibility of the crossover as a function of both the spatial pattern of spin flips and the structure of the interactions.



The rest of the paper is organised as follows. In \secref{sec:model}, we introduce the XY Hamiltonian acting on $N$ spin-$S$ particles which obey two long-range interaction patterns: the PWR2 and the power-law graphs. We also describe the set of initial states to be considered, which are built from a definite number of spin flips $k$ on a fully polarized state. 
We then study the quantum dynamics of the model for $S=1/2$ in \secref{sec:Quantum_model} by examining how bipartite and multipartite entanglement evolve as the number of initial spin flips increases. We observe that both quantities show complementary trends, and witness a crossover between collective order to scrambling in both their long-time average and temporal fluctuations. In \secref{sec:SC_model} we present a semi-classical analysis of the model in the large $S$ limit, relevant to atomic systems with cavity-mediated interactions \cite{hines2023}, where we characterize the stability of the collective dynamics through Lyapunov exponents.
We investigate why certain coupling geometries and initial states lead to highly chaotic behaviour by analyzing the eigenvalues and eigenvectors of the coupling matrix and their effect on the local stability of the system.
Then, in Sec. \ref{sec:spectral_analysis} we define a initial-state-dependent coupling graph and establish a connection between the spectral properties of its Laplacian and the robustness of collective dynamics in the system. Finally, in Sec. \ref{sec:conclusion} we present our conclusions and discuss avenues for future work. 

\section{Model}
\label{sec:model}
We study the one-dimensional spin-exchange Hamiltonian
\begin{align}
\label{eq:XY Hamiltonian}
H = \sum_{i,j} \chi_{ij} \left(S^{x}_{i}S^{x}_{j} + S^{y}_{i}S^{y}_{j}\right),
\end{align}
where $\chi_{ij}$ denotes the interaction strength between spins $i$ and $j$, as determined by the underlying coupling graph, $S_k^{\beta}$ represents a spin-$S$ operator on site $k$, and $\beta=x,y,z$. 
We consider two long-range spin models: the powers-of-two (PWR2) graph and the power-law graph, illustrated in Fig. \ref{fig:overview_fig} (b). In the PWR2 graph, two spins interact if their separation is an integer power of two, i.e.,
\begin{equation}
\chi_{ij} =
\begin{cases}
\chi_0, & \mathrm{if}\, |i-j| = 2^{m-1}, \\
0, & \mathrm{otherwise}.
\end{cases}
\label{eq:PWR2_graph}
\end{equation}
In the power-law graph, the interaction between spins $i$ and $j$ is given by $\chi_{ij} = |i-j|^{-\alpha}$. Throughout this work, we focus on the case $\alpha = 1$ with periodic boundary conditions.   

The choice of the Hamiltonian is motivated by its connection to the collective spin dynamics. In the uniform all to all case (e.g. $\alpha=0$), the XY Hamiltonian in \eqeqref{eq:XY Hamiltonian} reduces to the OAT Hamiltonian up to a constant. 
In previous works, it has been shown that that both the PWR2 and power law coupling geometries approximate the dynamics of the OAT \cite{Perlin_SS_2020,SK_metrology_2025}. The PWR2 graph is particularly striking because it achieves this with only a logarithmic number of couplings per site. 
In apparent contrast, earlier work \cite{Greg_tree_2019} showed that the same sparse graph Hamiltonian in \eqeqref{eq:XY Hamiltonian} can act as a fast scrambler when initialized from a disordered state. 
Taken together, these results suggest that the same family of long-range sparse Hamiltonians can support qualitatively distinct dynamical regimes depending on the choice of initial state. 
This motivates the central question of this work: is there a unified framework that explains the crossover between collective and scrambling regimes?


To explore this question, we consider a fiducial state where all spins are in the $\ket{+}$ state, and perform a series of $k$ spin flips in the $x$ basis (with $k=0,1,\ldots,N/2$) before letting the system evolve with $H$ for a time $t$. Although spins may, in principle, be flipped in any pattern, in Sec.~\ref{sec:Quantum_model} we focus on two representative protocols that capture the expected behavior most clearly. The first is one-sided flipping, where $k$ adjacent spins are flipped sequentially starting from a specific site of the one-dimensional ring. The second is alternating flipping, where spins are flipped on alternating sites; when $N/2$ spins are flipped in a system of $N$ spins, this protocol produces a Néel state in the $x$ basis, see Fig.~\ref{fig:overview_fig} (a). In Sec.~\ref{sec:SC_model} we will show that these two specific flipping patterns are particularly important in the context of translationally-invariant 1D chains by using graph-theoretic arguments. Nevertheless, for completeness we will also consider averages over randomly-chosen spin-flip patterns for some parts of our analysis. These different flipping patterns are illustrated in Fig.~\ref{fig:overview_fig} (a).

    \begin{figure}
    \centering
    \includegraphics[width=1.0\linewidth]{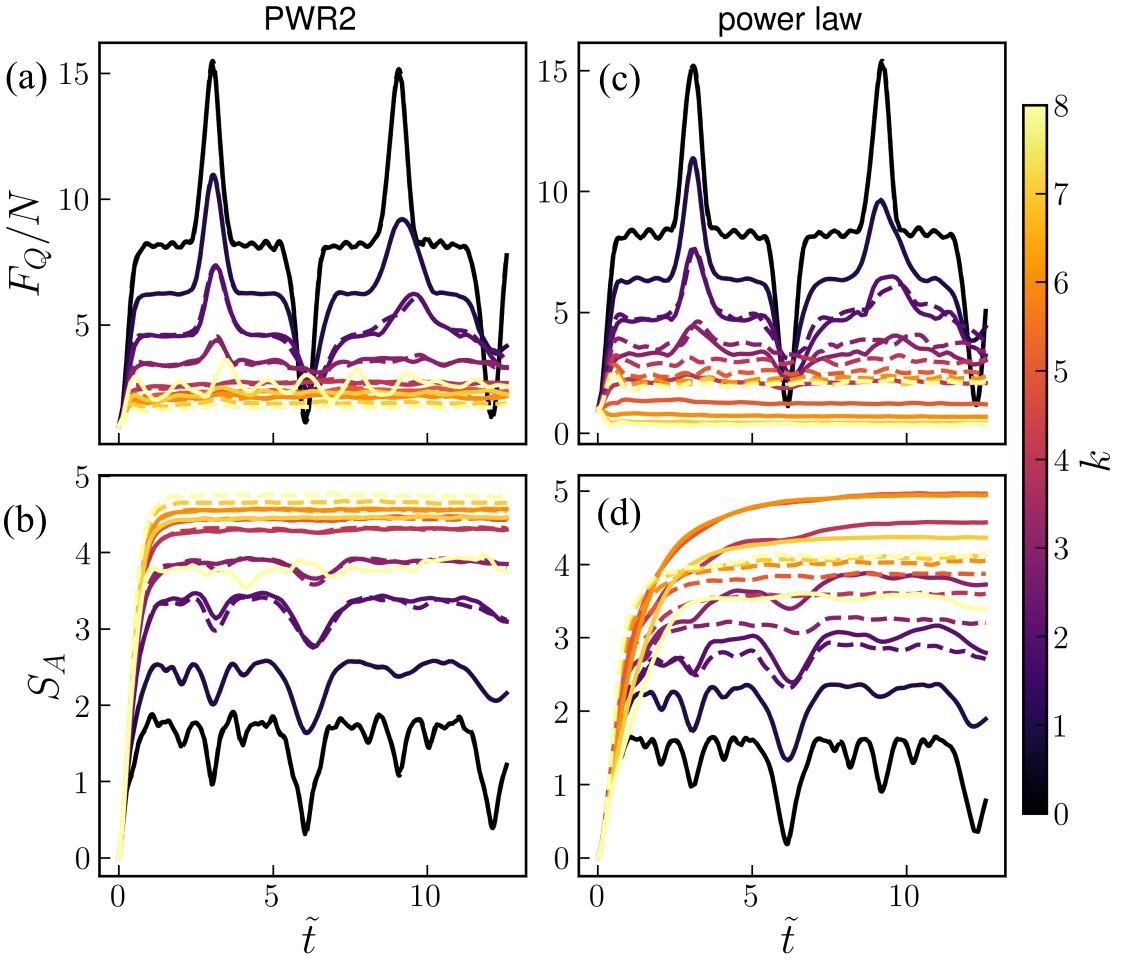}
    \caption{Evolution of the quantum Fisher information density $F_Q/N$ under the $XY$ Hamiltonian given in \eqeqref{eq:XY Hamiltonian} for (a) PWR2 periodic graph and (c) power law decaying graph with $\alpha=1.0$ for $N=16$ and different numbers of spin flips $k$. The time is normalized according to  \eqeqref{eq:time_norm}. Panels (b) and (d) show the bipartite entanglement entropy $S_A$  as a function of the normalized time for the corresponding graphs with $|A| = N/2$. The spins are initialized in an $x$-polarised state, and spin flips are performed in the $x$ basis. Solid and dashed lines correspond to alternating and one-sided spin-flip initial states, respectively. Any difference between the two curves is smaller than the line thickness. The number of spin flips $k$ is indicated by the colour bar.}
    \label{Fig:QFI_and_EE}
    \end{figure}
    
\section{Quantum Dynamics \label{sec:Quantum_model}}

We begin by exploring the changes in the quantum dynamics induced by performing spin flips in the initial state. In order to compare the evolution of different system sizes, we rescale time by the total coupling weight of the corresponding graph relative to the all-to-all graph \cite{SK_metrology_2025,Perlin_SS_2020}. We define the normalised time as
 \begin{align}
    \label{eq:time_norm}
    \tilde{t} = t\frac{\mathcal{E}_{G}}{\frac{N(N-1)}{2}} ~,
\end{align}
where $t$ is the physical time and $\mathcal{E}_{G}$ is the \emph{effective number of edges}, defined as the sum of the coupling weights,  
\begin{align}
    \mathcal{E}_{G}=\sum_{i<j}\chi_{ij}.
\end{align}
For the PWR2 graph, all non-zero couplings have unit weight, giving $\mathcal{E}_{G} = N\log_2(N) -\frac{N}{2}$. For the PL graph,
\begin{align}
\chi_{ij} =\frac{1}{r_{ij}^{\alpha}},
\qquad
r_{ij} = \min(|i-j|,N-|i-j|),
\end{align}
and hence $\mathcal{E}_{G} =\sum_{i<j}\frac{1}{r_{ij}^{\alpha}}$. 


To characterize how entanglement develops in time for different choices of initial spin flips, we study the bipartite entanglement entropy $S_A$, which quantifies quantum correlations between a subsystem $A$ and its complement. Specifically, we compute the von Neumann entanglement entropy,
\begin{equation} 
S_{A} = - \mathrm{Tr}[\rho_A \log{\rho_A}], 
\end{equation}
where $\rho_A = \mathrm{Tr}_{B}(\rho_{AB})$ is the reduced density matrix of subsystem $A$, which we take to be the result of a bipartition with $|A| = N/2$. In addition, we study the quantum Fisher information (QFI) $F_Q$, which captures the presence of multipartite entanglement \cite{PezzeSmerzi2014,PezzeSmerziOberthalerSchmiedTreutlein2018,BraunsteinCaves1994,Paris2009,Helstrom1969,TothApellaniz2014}. The QFI of a pure state $\ket{\psi}$, which determines the optimal precision for estimating a parameter $\theta$, is given by
\begin{align}
\label{Eq:QFI}
F_Q[\theta, J_\alpha] = 4 (\Delta J_\alpha)^2 ,
\end{align}
where $J_\alpha = \sum_i S_i^\alpha$ generates rotations about axis $\alpha = x,y,z$, and $(\Delta J_\alpha)^2 = \langle J_\alpha^2 \rangle - \langle J_\alpha \rangle^2$ is evaluated in the state $\ket{\psi}$. For our analysis we focus on $\alpha=x$, which is optimal for even $N$ in the one-axis-twisting Hamiltonian, corresponding to the all-to-all limit of Eq. (\ref{eq:XY Hamiltonian}). 

\begin{figure}
    \centering
    \includegraphics[width=1.0\linewidth]{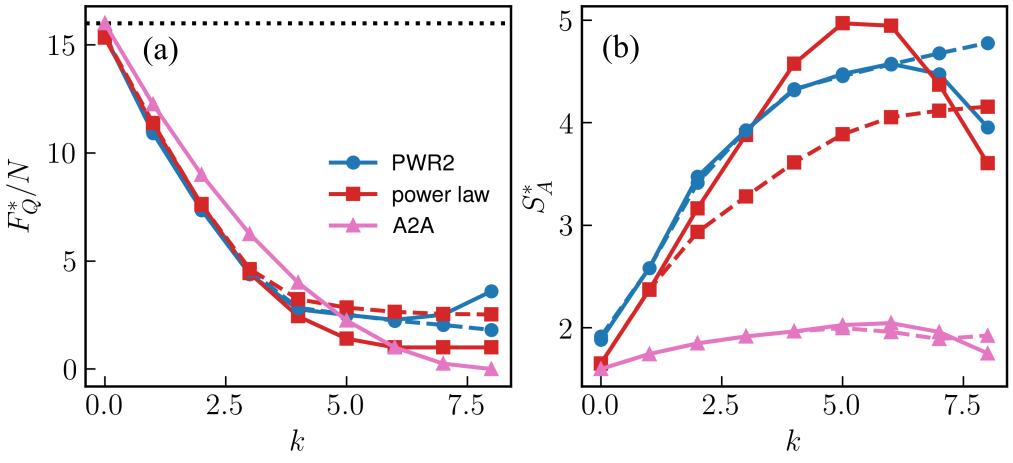}
    \caption{(a) Maximum quantum Fisher information per spin $F^{*}_{Q}/N$ extracted from the peak values in \figref{Fig:QFI_and_EE}(a) for the PWR2 graph and \figref{Fig:QFI_and_EE}(c) for the power-law decaying graph, shown as a function of the number of spin flips. The pink line indicates the corresponding scaling for the all-to-all (A2A) graph, and the horizontal black dotted line is a guide to the eye, showing the maximum value of $\frac{F_Q}{N}=N$. (b). Maximum bipartite entanglement entropy, $S_A^{*}$ extracted from \figref{Fig:QFI_and_EE}(b) for the PWR2 graph and (d) for the power-law decaying graph plotted versus the number of spin flips. Solid and dashed lines correspond to alternating and one-sided spin-flip initial states, respectively. 
    }
    \label{Fig:scaling_QFI_EE}
    \end{figure}

In \figref{Fig:QFI_and_EE}, we show the time evolution of both quantities for different initial states characterised by varying numbers of spin flips $k$, for a fixed system size of $N=16$. We observe that $S_A$ and $F_Q/N$ show complementary trends. 
For a number of spin flips $k\leq3$, the QFI exhibits qualitatively similar behavior for both coupling graphs (see panels \figref{Fig:QFI_and_EE} (a) and (c)), although the maximum values differ. 
As a function of time, the QFI initially increases, reaches a plateau, and attains its maximum around $\tilde{t} = \pi$. Beyond this point, it decreases again, displaying a periodic revival structure. 
This is a well-known behavior predicted by the collective one-axis-twisting Hamiltonian. However, as the number of spin flips increases, the QFI becomes suppressed, and its maximum value decreases with respect to the all-to-all case. The scaling of the maximum QFI is shown in \figref{Fig:scaling_QFI_EE} (a). For $k=0$, the maximum QFI $F^{*}_{Q}$ attains the value corresponding to the GHZ state. As $k$ increases, $F^{*}_{Q}$ reduces progressively for both the PWR2 and PL graphs. In the all-to-all graph (A2A), the scaling follows $F^{*}_{Q} \propto 4 \Big(\frac{N}{2}-k\Big)^2$, shown as the pink line for comparison. This quadratic scaling is expected, since in the A2A geometry every qubit contributes symmetrically to the collective spin, and removing $k$ excitations effectively reduces the minimal collective spin length from $\frac{N}{2}$ to $\frac{N}{2}-k$.  

By contrast, the entanglement entropy increases with the number of spin flips and eventually saturates  as shown in \figref{Fig:QFI_and_EE}(b,d). 
This indicates that the dynamics produces states which are no longer in permutationally symmetric manifold, and signals the redistribution of quantum correlations from collective multipartite entanglement to more localised bipartite entanglement.
The solid and dashed lines in \figref{Fig:QFI_and_EE} correspond to alternating and one-sided spin-flip configurations, respectively. Overall, the two protocols show similar dynamical behaviour, with only minor quantitative differences. The curves largely overlap on the scale shown, although small deviations in the QFI appear for larger numbers of spin flips. In Appendix~\ref{App:Avg_QFI_EE}, we show the corresponding signatures averaged over different choices of flipped sites for a fixed number of spin flips. 
In \figref{Fig:scaling_QFI_EE}(b), we plot the maximum bipartite entanglement entropy $S_A^{*}$ as a function of the number $k$ of spin flips. As $k$ increases, $S_A^{*}$ correspondingly increases. For the alternating spin-flip initial state, however, $S_A^{*}$ exhibits a dip at large $k$, a feature that is absent in the one-sided spin-flip case.

In \figref{Fig:QFI_and_EE}, we further observe a progressive suppression of late-time oscillations of both the QFI and bipartite entanglement entropy as the number of spin flips $k$ increases.
    \begin{figure}
    \centering
    \includegraphics[width=1.0\linewidth]{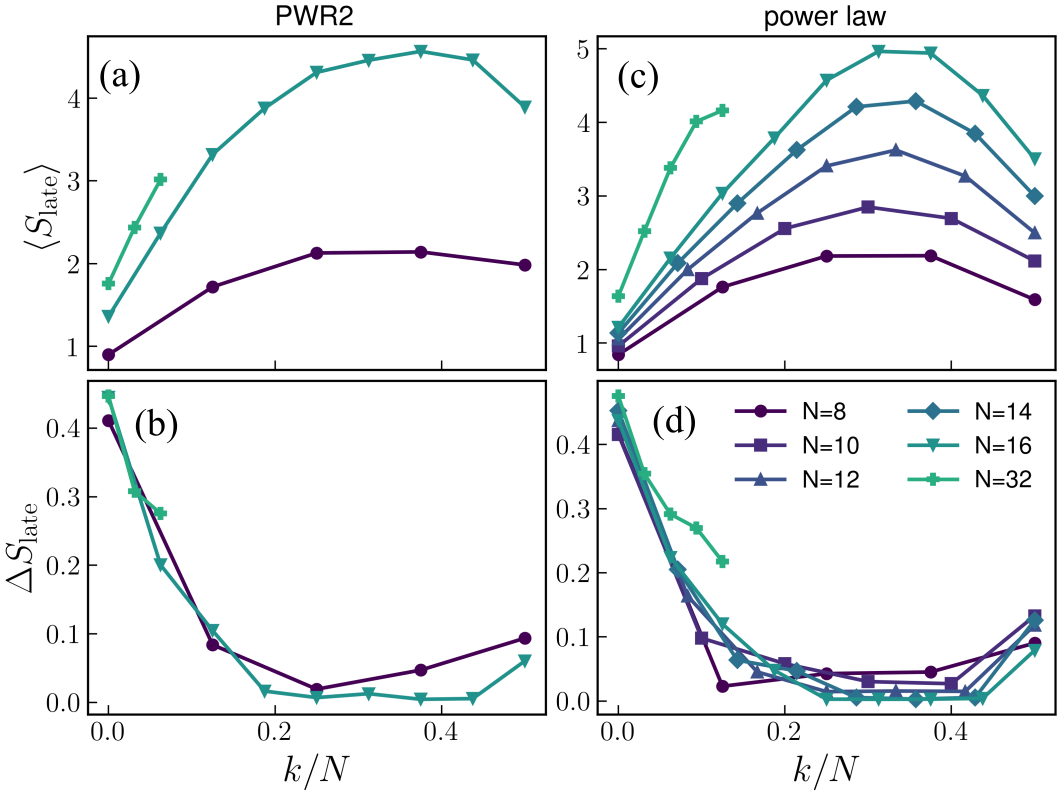}
    \caption{Average late time entanglement entropy $\langle S_{\mathrm{late}} \rangle$ under the dynamics of the $XY$ Hamiltonian in \eqeqref{eq:XY Hamiltonian}, where late times are defined as $\tilde{t} > 10$, plotted as a function of the fraction of flipped spins $k/N$ (with $k$ the number of spins flipped in the $x$-basis). Panels (a) and (b) correspond to the PWR2 periodic graph, and panels (c) and (d) to the PL coupling graph with with $\alpha =1.0$. Panels (a) and (c) show $\langle S_{\mathrm{late}} \rangle$, while panels (b) and (d) show the corresponding standard deviation of the late-time entanglement entropy \eqeqref{eq:standard deviation of S} characterising temporal fluctuations. We consider $N=8, 16, 32$ for PWR2 periodic graph and $N=8, 10, 12, 14, 16, 32$ for PL graph. For $N=32$ the number of data points is limited due to the rapid growth of entanglement, which restricts classical tensor-network simulations. 
    }
    \label{Fig:Revivals_and_late_EE}
    \end{figure}
To quantify this behaviour, we perform a detailed analysis of the time-dependence of the the bipartite entanglement entropy. 
Specifically, we compute the average late-time entanglement entropy $\langle S_{\mathrm{late}}\rangle$, where late times are defined as $\tilde{t} > 10$. 
The results are shown in \figref{Fig:Revivals_and_late_EE} (a) and (c) for PWR2 and PL graphs, respectively.
For both coupling graphs, $\langle S_{\mathrm{late}} \rangle$ generally increases with the fraction of flipped spins $k/N$, indicating enhanced late-time bipartite entanglement. 
At larger spin-flip fractions, however, this growth saturates and, for the alternating spin-flip configuration, exhibits again a noticeable drop near $k=N/2$. 
As with the results in Fig. \ref{Fig:scaling_QFI_EE}, this feature is absent for the one-sided spin-flip initial state as shown in Appendix~\ref{App:late_time_OS}. 

To further characterise the late time dynamics, we also calculate the temporal fluctuations of the entanglement entropy, 
    \begin{align}
    \label{eq:standard deviation of S}
        \Delta(S_A) = \mathrm{Stdv}(S_A(t) - \langle S_{\mathrm{late}}\rangle).
    \end{align}
These results are shown in \figref{Fig:Revivals_and_late_EE}(b,d). The fluctuations decrease rapidly as $k/N$ is increased, showing that the late-time entanglement dynamics become increasingly stationary, as expected for thermalizing dynamics. Again, near $k=N/2$ the alternating spin-flip initial state exhibits an increase in $\Delta(S_A)$. Since this configuration corresponds to a Néel state in the $x$ basis, this indicates that the Néel state retains stronger residual late-time oscillations compared with nearby spin-flip fractions.
Together, \figref{Fig:scaling_QFI_EE} and \figref{Fig:Revivals_and_late_EE} show that the Néel state displays qualitatively distinct behaviour compared with other spin-flip configurations.
This difference may be attributed to the strong period-2 spatial order of the initial state, which is invariant under translation by two sites and thus has additional discrete symmetry relative to the one-sided pattern. This more constrained structure can influence the unitary dynamics under the XY Hamiltonian, including the growth of entanglement and the extent of late-time revivals. More in-depth studies are left for future work.

We note that for small system sizes, we employed exact diagonalization, and for $N>16$, we use the time-dependent variational principle (TDVP) with matrix product states (MPS) \cite{WhiteFeiguin2004, DaleySchollwoeck2004, Schollwoeck2011, HaegemanVerstraete2016, FishmanStoudenmire2022, ITensor2022}. 
The results shown in the main text correspond to $N=32$ with bond dimension $D=512$, time step $\chi_0 dt = 10^{-2}$, and truncation error $\epsilon = 10^{-9}$. 
The limited number of data points for $N=32$ in Fig.~\ref{Fig:Revivals_and_late_EE} reflects the rapid growth of entanglement with increasing spin flips, making it challenging to capture the dynamics using MPS techniques. This suggests that the observed collective-to-scrambling crossover can also be regarded as a transition in the classical simulability of the dynamics.

To summarize, the results in this section indicate a noticeable crossover in the entanglement dynamics as number of spin flips in the initial state increases. The regime with $k\gtrsim0$ reveals extensive maximum QFI per particle, restricted entanglement entropy, and near-periodic dynamics with strong revivals. As $k$ increases towards $N/2$, the systems starts displaying sub-extensive QFI density, volumetric entanglement entropy, and equilibration to a stationary behaviour. 

\section{Semiclassical dynamics}
\label{sec:SC_model}

While the analysis in the previous section indicates that the dynamics shows signatures of chaotic behavior as the number of spin flips $k$ increases, a systematic exploration of scrambling requires calculation of out-of-time-ordered correlators (OTOCs). 
In quantum systems, the OTOC between local operators is defined as 
\begin{align}
    \label{eq:Quantum_OTOC}
    C(i,j,t) = \frac{1}{S^2} \Big \langle |[S_i^z(0), S_j^z(t)]|^2 \Big \rangle,
\end{align}
which measures how initially commuting operators fail to commute under time evolution.  
In a fast scrambler, OTOCs are expected to reach saturation on a timescale that grows logarithmically with system size, $\log(N)$.
\subsection{Classical Sensitivity}
To access large $N$ and long times we work in the semi-classical limit $S \gg 1$. 
This regime is directly relevant to cavity-mediated atomic systems, where large collective spins are naturally accessible \cite{Periwal_2021,qin_2019,Hung_2016,vaidya_2018,Mivehvar_2021,CavitySqueezing_Ian_2010,PNielson_molmer_squeezing}.
In this limit commutators map to Poisson brackets and the OTOC may be expressed as a classical sensitivity. Concretely, for an infinitesimal rotation $\phi_i$ of spin $i$ about $z$ axis the semi-classical OTOC (classical sensitivity) becomes
\begin{align}
\label{eq:OTOC}
    C_{cl}(i,j,t) = \frac{1}{S^2} \Bigg \langle \Bigg(\frac{dS_j^Z(t)}{d\phi_i}\Bigg)^2 \Bigg \rangle,
\end{align}
where $\langle .\rangle$ is taken over an ensemble of initial spin configurations. We note that this measure of sensitivity is related to the classical decorrelator, recently introduced in various many-body chaos studies \cite{kasim2025,ruidas2026}.
For a generic translationally invariant graph, we derive the $3N$ coupled differential equations starting from the Heisenberg Hamiltonian and solving the Heisenberg equations of motion in the large $S$ (mean-field) limit.
Details are provided in the Appendix \ref{app:classical_eoms}.
The resulting equations are
\begin{align}
\label{3n_eq_main}
\dot{S}^X_k &= \sum_{i} \chi_{i,k} S^Y_i S^Z_k, \nonumber \\
\dot{S}^Y_k &= -\sum_{i} \chi_{i,k} S^X_i S^Z_k, \nonumber \\
\dot{S}^Z_{k} &= \sum_{i}\chi_{i,k}
\left(S^X_iS^Y_k-S^Y_iS^X_k\right), 
\end{align}
where $\chi_{i,k}$ denotes the interaction coupling between sites $i$ and $k$.
These equations determine the classical trajectories used below to quantify the sensitivity to perturbations of the initial state.
We consider classical product states with $k$ flipped spins (relative to the fully $x-$polarized state ) and use \eqeqref{eq:OTOC} to compute the classical sensitivity. The resulting $C_{cl}$ depends on the number of spin flips and on the specific way they are introduced. Averaging over initial configurations is performed only when considering all possible placements of the $k$ flipped spins; for fixed flip patterns (alternative or one-sided), no ensemble average is taken. 
\begin{figure}
    \centering
    \includegraphics[width=1.0\linewidth]{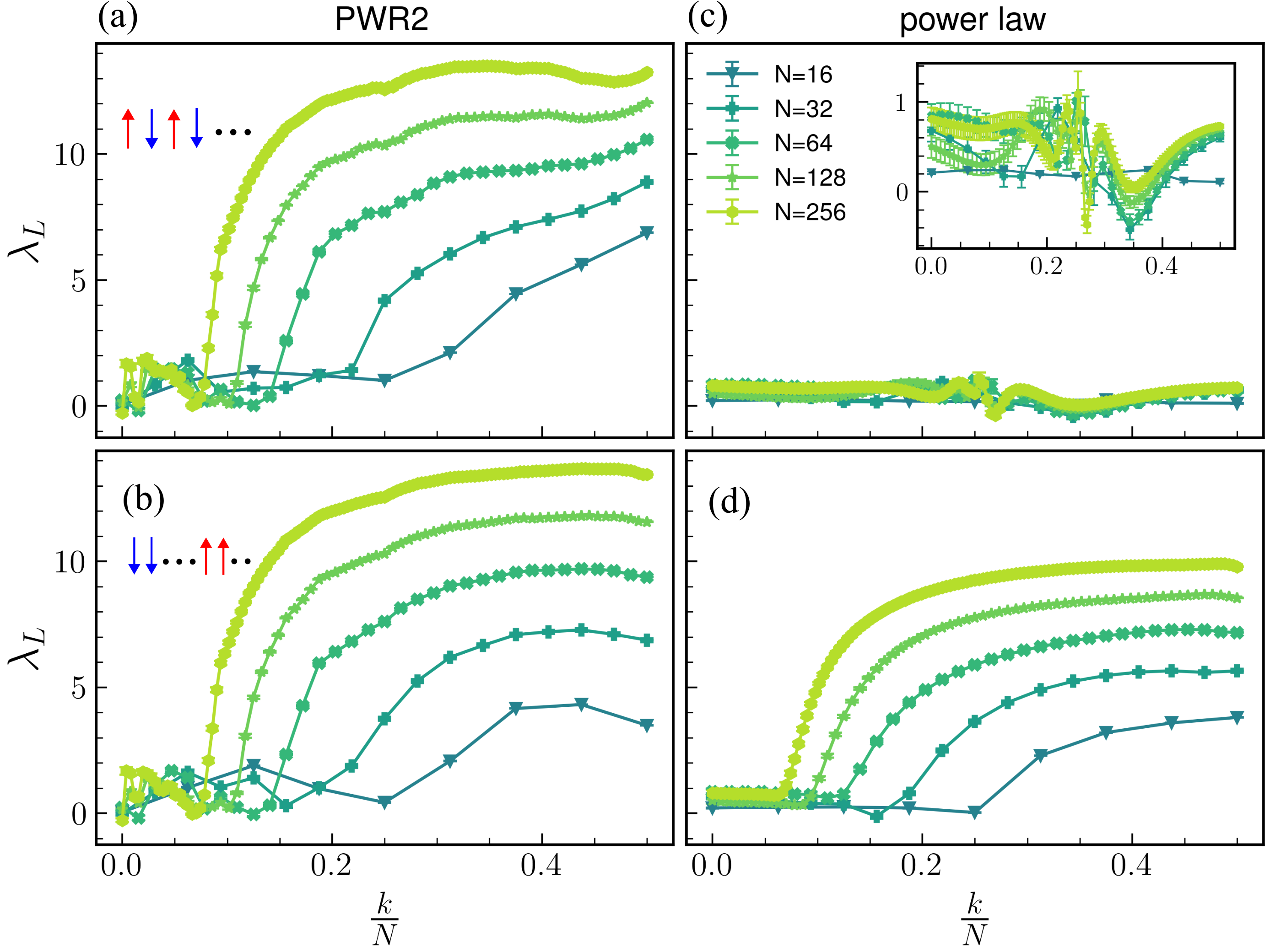}
    \caption{Lyapunov exponent extracted from the semi-classical OTOC \eqeqref{eq:OTOC} as a function of the fraction of flipped spins $k/N$. Panels (a) and (b) correspond to PWR2 periodic graphs, and panels (c) and (d) to power law decaying graph with $\alpha = 1.0$, for $N = 16, 32, 64, 128, 256$. We consider two spin-flipping protocols: alternating flips, schematically illustrated in panel (a), and one-sided flips, schematically illustrated in panel (b). The inset in (c) shows a zoomed-in view of the corresponding data.Error bars indicate the standard deviation obtained from fitting the exponential growth regime. 
    }
    \label{fig:Lyapunov_exp}
\end{figure}
\\
Still, for a fixed initial state, the $C_{cl}$ depends on $i$ (the perturbed site) and $j$ (the probe site).
At early times, $\chi_0 t \lesssim 1$, the classical sensitivity exhibits a polynomial growth, $C_{cl}(t) \propto (\chi_0 t)^{2r_{ij}}$ where $r_{ij}$ is the graph distance between sites $i$ and $j$ \cite{Greg_tree_2019}. Then, $C_{cl}$ crosses over to an exponential regime for $\chi_0 t \gtrsim 1$ irrespective of $i$ and $j$, from which the corresponding Lyapunov exponent $\lambda_L$ can be extracted. In Appendix \ref{app:lyapunov_calc} we provide further details about this procedure.

In \figref{fig:Lyapunov_exp}, we plot $\lambda_L$ as a function of the normalized number of spin flips, $k/N$, for the PWR2 graph [panels (a) and (b)] and for the power-law decaying graph [panels (c) and (d)]. We consider two distinct initial configurations: alternating spin flips (top panels) and one-sided spin flips (bottom panels). 
For the PWR2 graph, we observe a clear increase in Lyapunov exponent with increasing $k/N$, reflecting regions of scrambling. 
Moreover, this rise shifts toward smaller values of $k/N$ as the system size increases, indicating that larger systems scramble more efficiently.
This qualitative behaviour is robust across both types of initial configurations.
In contrast, for the power-law decaying graph we find less generic behavior, with the Lyapunov exponent remaining negligible for the alternating configuration (see \figref{fig:Lyapunov_exp} (c) and \figref{fig:Lyapunov_exp} (c) inset), but growing with $k/N$ for one-sided spin flips.
This is reminiscent of the quantum results in Sec. \ref{sec:Quantum_model}, where alternating spin flips were also found to suppress chaos for both PWR2 and PL coupling graphs.
Finally, we note that for spin flips $k=N/2$ and $k=N/4$, we computed the time $t^{*}$ at which $C_{cl}(t^{*})=1$ in order to examine the scaling of $\lambda_Lt^{*}$ with system size $N$. This is discussed in Appendix \ref{app:Scaling_lyp}. 

\begin{figure}
    \centering
    \includegraphics[width=1.00\linewidth]{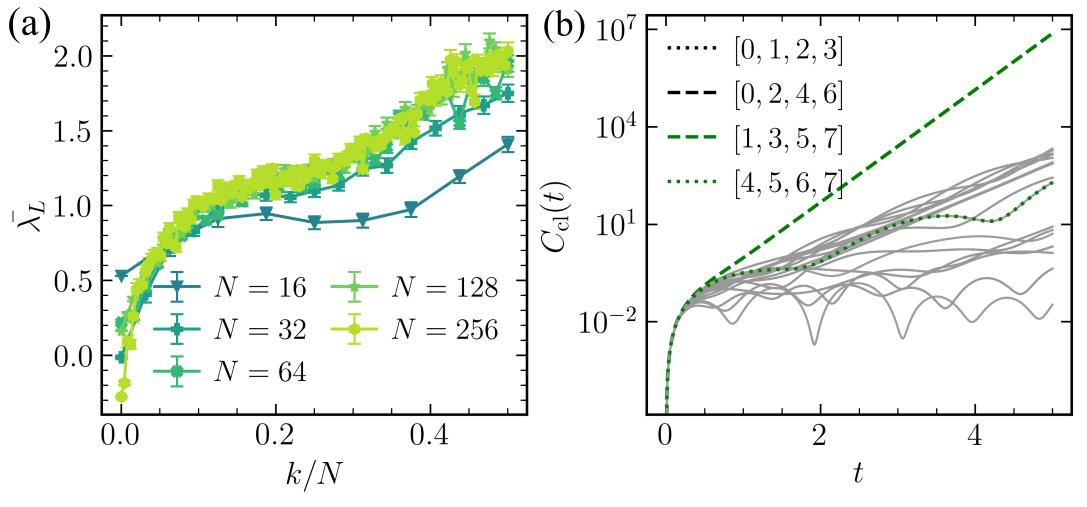}
    \caption{(a) Averaged Lyapunov exponent $\bar{\lambda}_L$ as a function of the fraction of flipped spins $k/N$ for different system sizes $N$. The error bars reflect the statistical uncertainty across spin-flip configurations, while the overall values of $\bar{\lambda}_L$ remains small. (b) To understand this, we plot all 70 trajectories for $N=8$ with $k = N/2 = 4$ flipped spins. The alternating spin configuration (dashed black and green curves) exhibits the largest growth in classical sensitivity $C_{\mathrm{cl}}$, defined in \eqeqref{eq:OTOC}. The other trajectories are shown as grey lines. This behaviour persists for larger system sizes.}
    \label{fig:Lyapunov_exp_averaged}
\end{figure}

For the remainder of this section, we focus on the PWR2 graph for which we expect more generic behavior. To further support our analysis, we average $\lambda_L$ over an ensemble of initial states. For a system of $N$ spins with $k$ flipped spins, there are $\binom{N}{k}$ distinct spin-flip configurations. When $\binom{N}{k} \leq 200$, we average over all such configurations, otherwise we sample 200 configurations uniformly at random. The resulting averaged Lyapunov exponent $\bar{\lambda_L}$  is shown in \figref{fig:Lyapunov_exp_averaged} for different system sizes. 

In Fig.~\ref{fig:Lyapunov_exp_averaged} (a), we observe that the averaged Lyapunov exponent $\bar{\lambda_L}$ reaches maximum values $\sim 2$, well below what is achievable for one-sided and alternating spin flips. The error bars are obtained by taking the standard deviation and dividing it by the square root of the number of samples, thereby reflecting the statistical spread over different spin-flip configurations. To understand why $\bar{\lambda_L}$ remains relatively small, we consider a system of size $N=8$ with $k = N/2$, which yields 70 distinct spin-flip configurations. The classical sensitivity, $C_{\mathrm{cl}}$, for all configurations with four flipped spins is shown in Fig.~\ref{fig:Lyapunov_exp_averaged} (b) as a function of time $t$. Two representative flipping patterns are highlighted in the legend: alternating (dashed green and black) and one-sided (dotted green and black). We observe a significantly faster increase in $C_{\mathrm{cl}}$ for alternating flips compared to other configurations. The strong enhancement of $C_{\mathrm{cl}}$ for alternating configurations raises the question of why these patterns behave differently from others. In the next section, we address this question for the PWR2 graph and for general periodic coupling graphs.

\subsection{Linearised equations of motion}
We consider coupling graphs with periodic boundary conditions, for which the adjacency matrices are circulant. A circulant matrix is fully specified by its first row, as each subsequent row obtained by a cyclic shift from the previous one. 
For an $N \times N$ circulant matrix, the eigenvectors are the discrete Fourier modes
\begin{align}
\label{eq:eigenvector_cm}
    f_q = \frac{1}{\sqrt{N}} \left( 1, e^{\frac{2\pi i q}{N}}, e^{\frac{2\pi i q \cdot 2}{N}}, \dots, e^{\frac{2\pi i q (N-1)}{N}} \right)^T,
\end{align}
with corresponding eigenvalues
\begin{align}
    \lambda_q = \sum_{r=0}^{N-1} a_r \, e^{-\frac{2 \pi i q r}{N}},
\end{align}
where $a_r$ are the entries of the first row of the matrix.

Among these modes, two configurations play a important role: the uniform mode $q=0$, corresponding to $(1,1,\dots,1)$, and the staggered (alternating) mode $q = N/2$ (for even $N$), corresponding to $(1,-1,1,-1,\dots)$. These modes also correspond to initial states of interest in our system (E.g. fully polarized and alternating, respectively).

To illustrate how the eigenmodes of the adjacency matrix govern the dynamics of the classical sensitivity, we first note that every initial spin configuration satisfying $S_j^Z=0$, $\forall j$ (i.e. with all spins lying on the equator) corresponds to a fixed point of the equations of motion (see Appendix \ref{app:classical_eoms}). We may therefore linearize the dynamics about any such configuration. 
The resulting linearized equations of motion can be written in terms of the circulant coupling matrix $\chi$. For the alternating configuration, which is itself an eigenvector of $\chi$, the linearized operator remains circulant. As a result, perturbations can be decomposed into Fourier modes that evolve independently. The equations of motion (derived in Appendix~\ref{app:linear_eoms}) then reduce to
\begin{align}
    \ddot{y}_q = \lambda_{N/2}(\lambda_q - \lambda_{N/2}) \, y_q,
\end{align}
where $\lambda_{N/2}$ is the eigenvalue associated with the alternating background.
The growth of perturbations is therefore controlled by the sign of $\lambda_{N/2}(\lambda_q - \lambda_{N/2})$. In particular, exponential growth and hence enhanced classical sensitivity (or OTOC growth) occurs when
$\lambda_{N/2}(\lambda_q - \lambda_{N/2}) > 0$.

\begin{figure}
    \centering
    \includegraphics[width=1.0\linewidth]{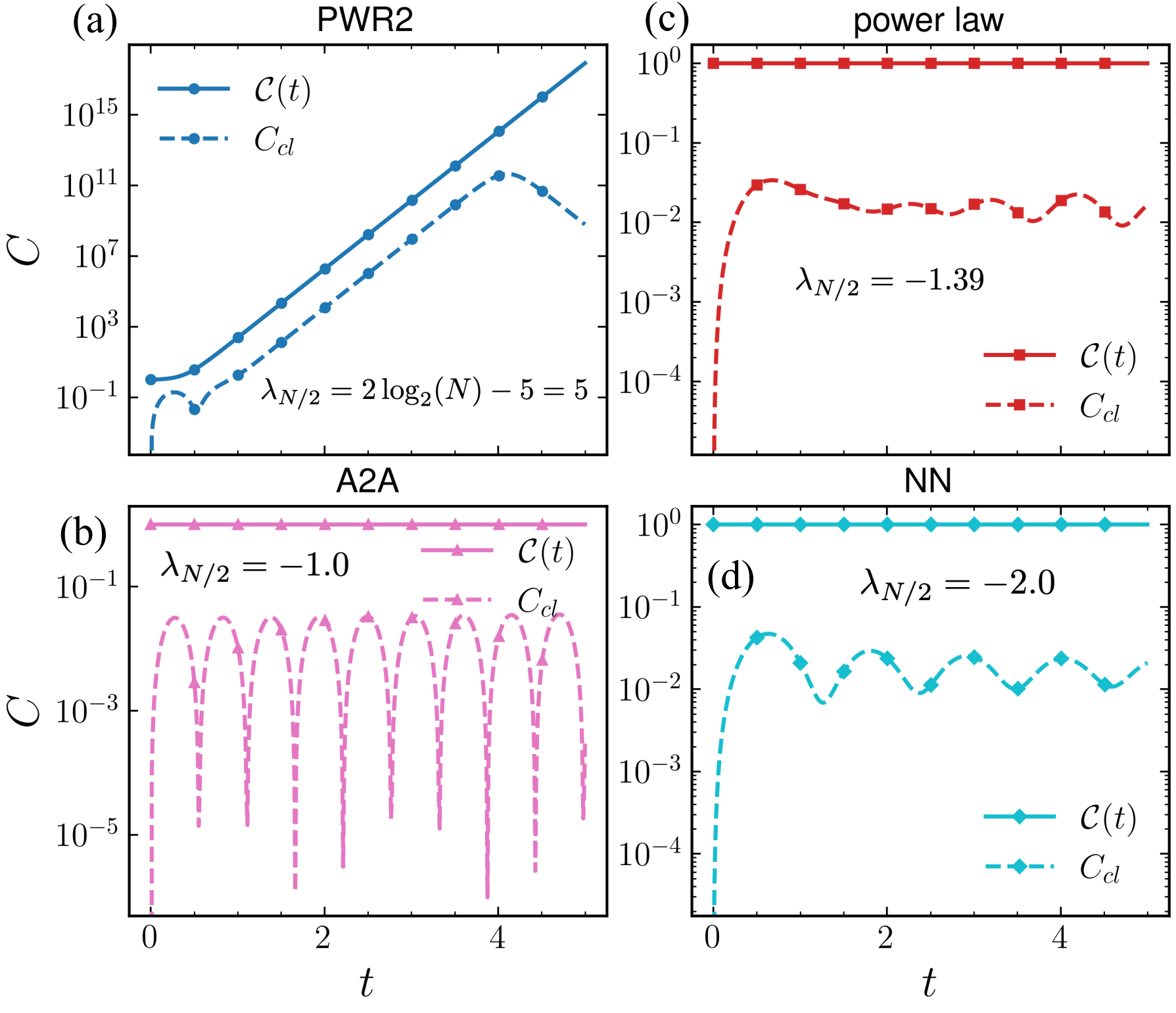}
    \caption{Classical sensitivity $C_{\mathrm{cl}}$ (dashed), defined in  \eqeqref{eq:OTOC} and $\mathcal{C}(t)$ (solid), defined in \eqeqref{eq:recipe_otoc} as functions of time $t$ for different periodic coupling matrices: (a) PWR2, (b) power-law decay with $\alpha = 1.0$, (c) all-to-all (A2A), and (d) nearest-neighbor graphs. Notably, $\lambda_{N/2} > 0$ only for the PWR2 geometry among the coupling graphs considered.}
    \label{fig:Circulant_otoc}
\end{figure}

In the following, we compare this mechanism across four representative coupling matrices relevant to the Hamiltonian under consideration. To do so, we compute an approximation to the classical sensitivity, which we denote by $\mathcal{C}(t)$.
To define the initial perturbation, we flip the spin at a given site while leaving all other spins initially unchanged.
The corresponding mode coefficients $c_q$ are then obtained by projecting this perturbation onto the eigenvectors of the coupling matrix $\chi$. At time $t$, each mode evolves independently with amplitude $y_q(t) \sim c_q e^{\Gamma_q t}$, where 
$\Gamma_q(K=\frac{N}{2}) = \sqrt{\lambda_{N/2}(\lambda_q - \lambda_{N/2})}$, where $K$ determines the type of background, in this case alternating spin flips.  
Defining the perturbation vector as
\begin{align}
  \delta\mathbf{y}(t)=\big(\delta y_1(t),\ldots,\delta y_N(t)\big)^{T},  
\end{align}
$\mathcal{C}(t) \propto \|\delta\mathbf{y}(t)\|^2$. The total sensitivity is obtained by summing over modes,
\begin{align}
\label{eq:recipe_otoc}
  \mathcal{C}(t) \sim \sum_q |c_q|^2 e^{2\Gamma_q(K) t}.  
\end{align}
We use this expression to compute the early-time behavior of $\mathcal{C}(t)$ for different coupling geometries. 
The results are shown in \figref{fig:Circulant_otoc} for $N=32$.
The dashed curves correspond to the classical sensitivity $C_{\mathrm{cl}}$, defined in \eqeqref{eq:OTOC}, while the solid curves show the prediction from $\mathcal{C}(t)$. 
For the PWR2 graph, we find excellent agreement in the slope at early times, indicating that the growth of classical sensitivity is well captured by the eigenvalue spectrum of the circulant coupling matrix for the alternating spin configuration with $N/2$ spin flips.
This framework also provides a convenient way to compare different coupling geometries. Notably, the PWR2 graph is the only case among those considered for which $\lambda_{N/2} > 0$ (see \figref{fig:Circulant_otoc}(a)), whereas for power-law, all-to-all, and nearest-neighbour couplings (panels (b)–(d)), $\lambda_{N/2} < 0$.
This distinction is crucial: it implies that, among these models, only the PWR2 geometry supports an instability of the alternating configuration, and hence exhibits a crossover from collective spin behaviour to chaotic dynamics under alternating spin flips. 

\begin{figure}
    \centering
    \includegraphics[width=1.0\linewidth]{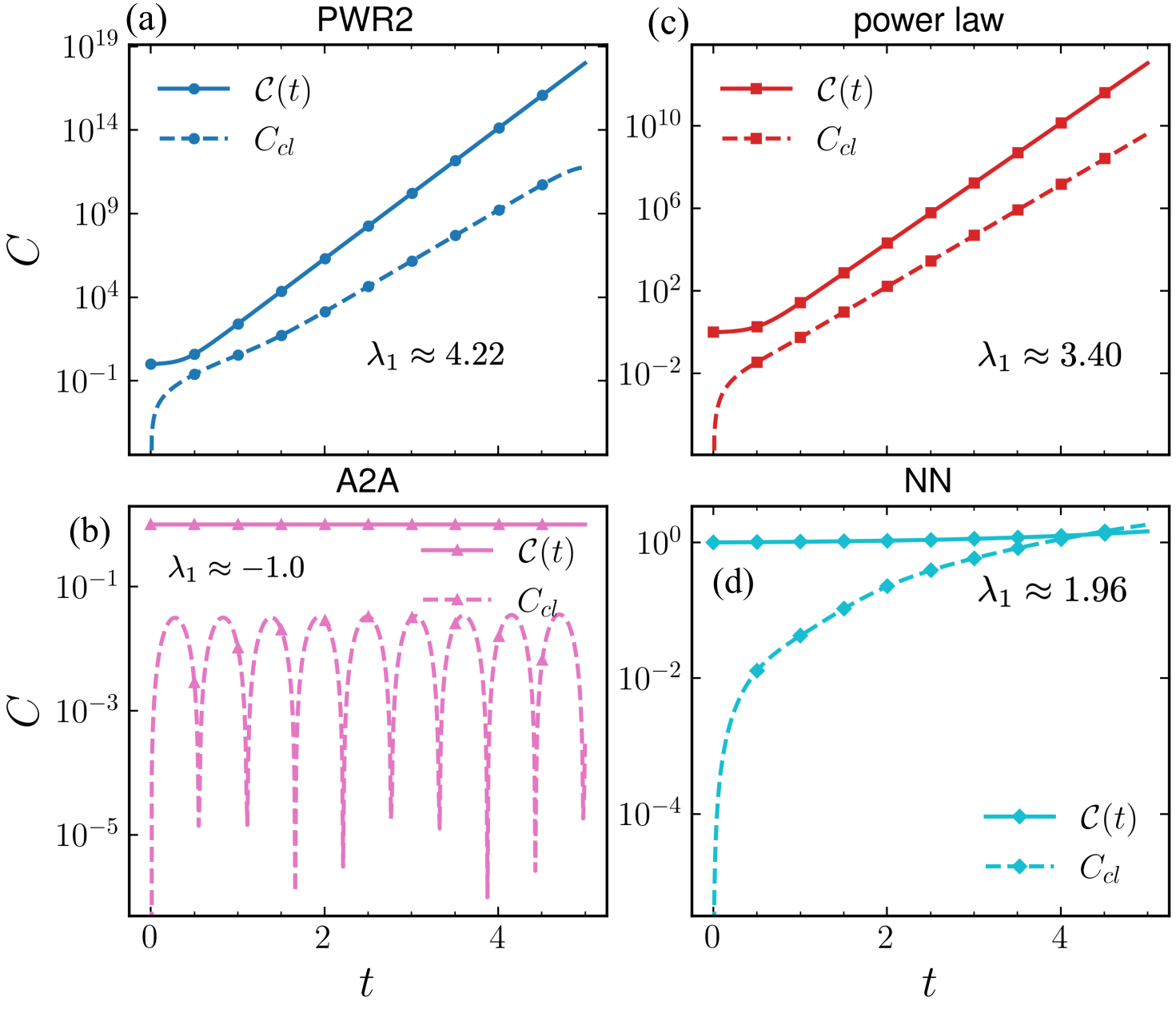}
    \caption{Classical sensitivity $C_{\mathrm{cl}}$ (dashed), defined in  \eqeqref{eq:OTOC} and $\mathcal{C}(t)$ (solid), defined in \eqeqref{eq:recipe_otoc} as functions of time $t$ for  different periodic coupling matrices: (a) PWR2, (b) PL coupling graph with $\alpha = 1.0$, (c) all-to-all (A2A), and (d) nearest-neighbor graphs. These results are for the one-sided flipped background and are calculated approximately using $\lambda_1$, which is positive for all coupling geometries except A2A. This shows that the power-law-decay graph exhibits increasing $\mathcal{C}(t)$ for one sided spin flips as the initial state.}
    \label{fig:RECIPE_OTOC_OS}
\end{figure}

It is interesting to note that the eigenvectors of the circulant matrix given in \eqeqref{eq:eigenvector_cm} are independent of the coupling matrix. Thus, once the overlap of an initial state with these modes is known, different initial states can be compared using the same Fourier basis. 
The alternating state and the no-spin-flip (fully polarized) state are exact eigenvectors of the circulant matrix (we show the analysis for the latter in Appendix \ref{app:eigenvectors_Uniform_OS}).  However, the one-sided spin-flip state is not an eigenvector, but its overlap with the Fourier modes can still be computed. 
This overlap is largest for $q=1$ and $q=N-1$. 
One can also show that the one-sided spin-flip state $(-1, -1, -1, \cdots 1,1)$ has support only on odd $q$ (See appendix \ref{app:eigenvectors_Uniform_OS}).
We therefore approximate $\Gamma_q^{\mathrm{eff}}(K=1)\simeq\sqrt{\lambda_1(\lambda_q - \lambda_1)}$, where (K=1) denotes the one-sided spin-flip background. We use this approximation to calculate $\mathcal{C}(t)$, which we show in \figref{fig:RECIPE_OTOC_OS} for $N=32$ for different coupling geometries. The dashed curves correspond to the classical sensitivity $C_{\mathrm{cl}}$, defined in \eqeqref{eq:OTOC}, while the solid curves show the prediction from $\mathcal{C}(t)$ defined in \eqeqref{eq:recipe_otoc}. 
We see that, in addition to PWR2, the PL coupling graph also shows an increase in $\mathcal{C}(t)$, as shown in \figref{fig:RECIPE_OTOC_OS} (a) and (c).  
This supports the results in \figref{fig:Lyapunov_exp}, where we observed a rise in the Lyapunov exponent for one-sided spin flips in the PL graph, but no rise for alternating spin flips. We also see that the value of $\lambda_1$ is positive for all coupling graphs except A2A, as shown in \figref{fig:RECIPE_OTOC_OS}(b). 
The nearest neighbour coupling graph has a smaller positive $\lambda_1$ (see \figref{fig:RECIPE_OTOC_OS} (d)) than PWR2 and power law decay, reflecting a weaker growth in $\mathcal{C}(t)$. 
For the one-sided spin-flip state, this prediction can overestimate the growth because the state is not an eigenvector of the circulant matrix. In the approximation above, only the contribution from $\lambda_1$ is retained and the other harmonics like $q=3,5$ are neglected. These modes may grow slower. Nevertheless, using this argument, we were able to show that power law decay coupling graph can give rise to an exponentially increasing $\mathcal{C}(t)$. 

\begin{figure}
    \centering
    \includegraphics[width=1.0\linewidth]{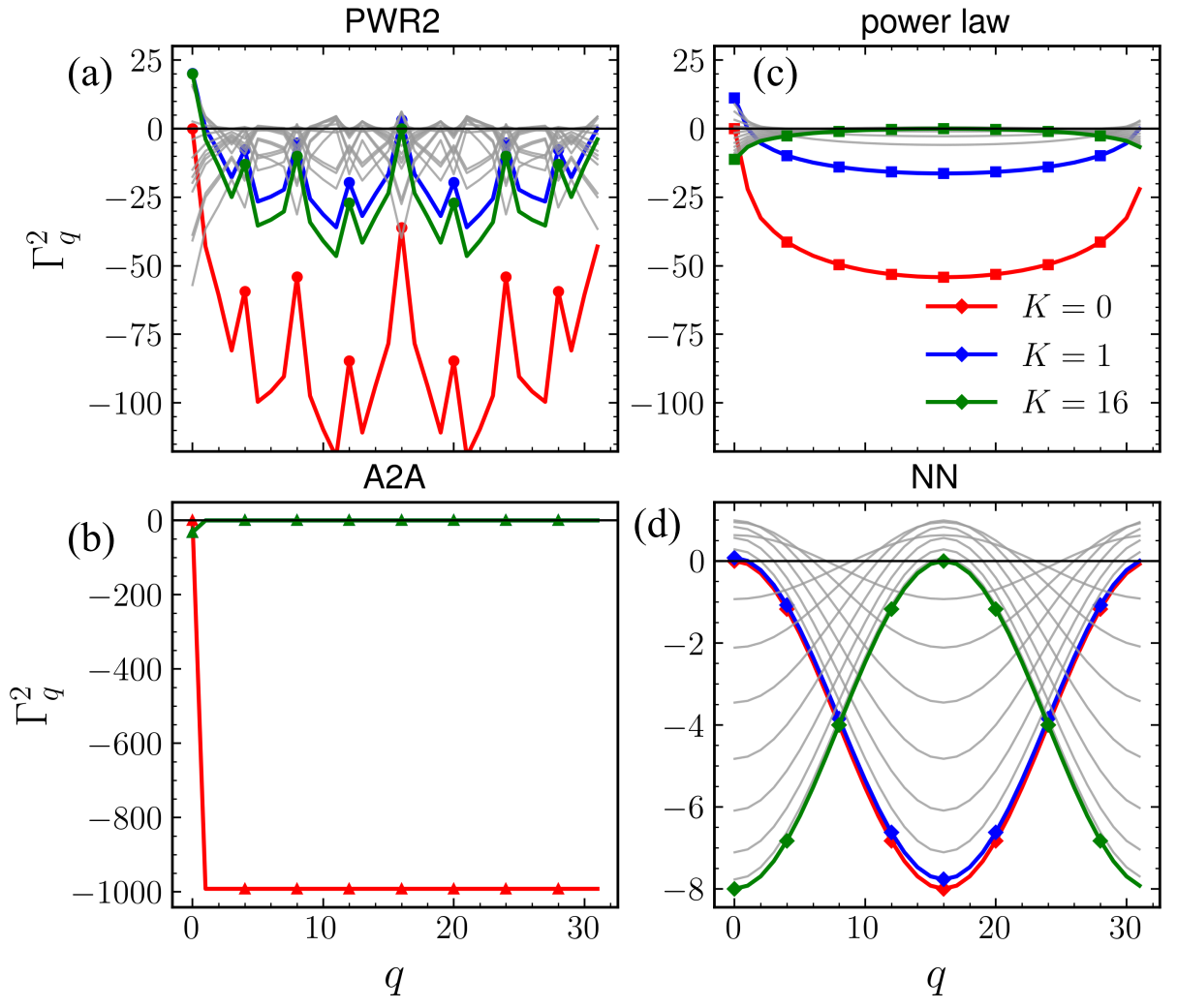}
    \caption{$\Gamma_q^2(K)$ deined in \eqeqref{eq:Gamma_K} as a function of modes $q$ for different $K$ and for different periodic coupling geometries: (a) PWR2, (b) PL coupling graph with $\alpha = 1.0$, (c) all-to-all (A2A), and (d) nearest-neighbor graphs. The red, blue, and green lines correspond to $K=0, 1, N/2$ respectively. The grey lines correspond to other values of $K \in \{2,3, \cdots \frac{N}{2}-1\}$. The horizontal black line is a guide to the eye plotted at $\Gamma_q^2=0$. For clarity of the figure, only a subset of the computed data points are shown.}
    \label{fig:gamma_score}
\end{figure}

We stress that the analysis presented above allows us to predict the onset of exponential growth of the classical sensitivity without actually simulating the dynamics, but rather by exploiting the spectral properties of the adjacency matrix alone. 
Given this tool, it is natural to ask what happens for more general initial states. A similar analysis can be used to compute the overlap with the Fourier modes, identify the dominant mode, and thereby obtain an approximate $\mathcal{C}(t)$. Equivalently, we can plot $\Gamma_q^2(K)$ for different values of $K$, where $K$ determines the type of background: $K=N/2$ for alternating spin flipped background, $K=0$ for uniform unflipped (polarized) background. 
One can hence compute 
\begin{align}
\label{eq:Gamma_K}
    \Gamma_q^2(K) = \lambda_K(\lambda_q - \lambda_K),
\end{align}
and study how this changes with $K$. 
This is shown for all the different coupling graphs in \figref{fig:gamma_score}. The figure highlights $K=0,1,N/2$; other values of $K$ are shown as grey. 
We see that the PWR2 coupling graph has the largest value of $\Gamma_q^2$ at $q=0$, as shown in \figref{fig:gamma_score}(a). 
The PL coupling graph has a positive $\Gamma_q^2$ for $K=1$, which is representative of the approximately one-sided flipped background, as shown in \figref{fig:gamma_score}(b). 
As discussed previously and shown in \figref{fig:Circulant_otoc}, $K=N/2$, which corresponds to the alternating background, gives a positive $\Gamma_q^2$ only for the PWR2 coupling graph. The A2A coupling graph has $\Gamma_q^2\leq0$ for all $K$, as shown in \figref{fig:gamma_score}(c), while the nearest neighbour coupling graph has a few grey lines above $\Gamma_q^2=0$; however, for the backgrounds we consider, $\Gamma_q^2\leq 0$. as shown \figref{fig:gamma_score}(d). 
\section{Spectral analysis}
\label{sec:spectral_analysis}

\subsection{Laplacian gap and collective states}
In this section we return to the quantum model ($S=1/2$) and analyze the collective-to-scrambling crossover by studying the spectral gap between the collective subspace and the rest of the Hilbert space in long-range interacting Hamiltonians. We recall that \eqeqref{eq:XY Hamiltonian} can be written as \cite{Perlin_SS_2020,SK_metrology_2025}
\begin{align}
    H = H_{\mathrm{gOAT}} + V_{\mathrm{pert}},
\end{align}
where
\begin{align}
    H_{\mathrm{gOAT}} &= \chi_0 J_z^2 - \sum_{ij} \chi_{ij}\vec{S_i}.\vec{S_j}\hspace{0.2cm} \mathrm{and} \\
    V_{\mathrm{pert}} &= -\sum_{ij}(\chi_0-\chi_{ij})S_i^zS_j^z.
    \label{eq:hami_goat}
\end{align}
Here, $J_z=\sum_{i}S_i^z$ is the collective spin along the $z$ direction and $H_{\mathrm{gOAT}}$ is the generalized one-axis-twisting Hamiltonian, which includes an isotropic Heisenberg term. This Hamiltonian commutes with the total spin $J^2 = J_x^2 + J_y^2 + J_z^2$, i.e. $[H_{\mathrm{gOAT}},J^2]=0$, and thus preserves collective states. However, $V_{\mathrm{pert}}$ acts as a perturbation that drives the state out of the permutationally symmetric manifold. The effect of the perturbation depends on the size of the spectral gap  $\Delta$ between the collective ($J=N/2$) eigenstates of $H_{\mathrm{gOAT}}$ and the rest of the states. This gap is completely determined by the Heisenberg coupling and provides a measure of how well collective states are protected from permutation-breaking interactions (see also \cite{lerose2025theory,panniergunther2025longrange}). This approach has been used in previous works to analyze the onset of scalable spin squeezing \cite{Perlin_SS_2020} and quantum control \cite{panniergunther2026}. In our previous work, we showed that sparse coupling graphs like the PWR2 model lead to a gap that is constant with system size \cite{SK_metrology_2025}.


For a given coupling graph $G = \{\chi_{ij}\}$, the minimum gap $\Delta$ can be computed from Laplacian of the graph. The graph Laplacian is defined as $L = D - A$, where $A$ is the adjacency matrix and $D$ is the degree matrix, $D_{ij}=\delta_{ij}\sum_k A_{ik}$. When the edges of the graph have positive weight, $A_{ij}\geq0$, $L$ is positive semidefinite, its minimum eigenvalue is $l_0=0$, and the gap is determined by the smallest positive eigenvalue $l_1=\Delta$. In this case, there is a one-to-one correspondence between the dynamical robustness of initial collective states and the mixing time of a discrete (classical) random walk on the graph $G$, as both quantities are determined by $\Delta^{-1}$ \cite{chung1996spectral}.

\subsection{Extension to non-collective states}

The construction in the preceding section applies to dynamics of collective states or, more generally, eigenstates of $J^2$. The initial states we consider here (see Fig. \ref{fig:overview_fig}) do not possess this property except for $k=0$. To extend this framework to such cases, consider an initial state with $k$ spin flips written as
\begin{align}
    \ket{\Psi^{\boldsymbol{\alpha}}_k(0)} &= \mathcal{Z}_k^{\boldsymbol{\alpha}} \ket{+}^{\otimes N}, \hspace{0.2cm} \mathrm{where} \\
     \mathcal{Z}_k^{\boldsymbol{\alpha}} &= Z_{\alpha_1}Z_{\alpha_2}..Z_{\alpha_k},
\end{align}
and $\boldsymbol{\alpha} = (\alpha_1,..\alpha_k)$ specifies the location of the spin flips. The time evolution of this state is then
\begin{align}
    \ket{\Psi^{\boldsymbol{\alpha}}_k(t)} &= e^{-iHt} \mathcal{Z}_k^{\boldsymbol{\alpha}}\ket{+}^{\otimes N} = \mathcal{Z}_k^{\boldsymbol{\alpha}}e^{-iH^{\boldsymbol{\alpha}}_kt} \ket{+}^{\otimes N},
    \label{eq:evol_flips}
\end{align}
where we used that $\mathcal{Z}_k^{\boldsymbol{\alpha}}$ commutes with $H$ up to a similarity transformation. This defines an effective Hamiltonian 
\begin{equation}
    H_k^{\boldsymbol{\alpha}} = \mathcal{Z}_k^{\boldsymbol{\alpha}} H \mathcal{Z}_k^{\boldsymbol{\alpha}}=\sum_{i,j} \chi_{ij}^{\boldsymbol{\alpha}} \left(S^{x}_{i}S^{x}_{j} + S^{y}_{i}S^{y}_{j}\right)
    \label{eq:transf_hami}
\end{equation}
which preserves the same connectivity of the original graph but where some of the edges have a negative sign (see Appendix \ref{app:spin_flips_gap}) depending on the initial state flipping pattern, labelled by $\boldsymbol{\alpha}$. Nevertheless, a key property remains unchanged: the associated graph Laplacian continues to represent the Hamiltonian of an isotropic Heisenberg model. 

Note from \eqeqref{eq:evol_flips} that, thanks to this transformation, the initial state being evolved by the transformed Hamiltonian is the collective fully polarized state. However, an additional flipping operation is required at the end of the protocol to map back to the original dynamics. Since such local operations will not affect the entanglement structure of the state, we disregard its effect in our analysis. Finally, because the transformed graph $G_{\boldsymbol{\alpha}}=\{\chi_{ij}^{\boldsymbol{\alpha}}\}$ is signed, its Laplacian, while still symmetric, is no longer positive semidefinite (and thus cannot be seen as the generator of a random walk). Nonetheless, the eigenvalues associated with the collective subspace is still $l_0=0$. We define $\Delta$ in this general case as the smallest eigenvalue difference with respect to this value.

\begin{figure}
    \centering
    \includegraphics[width=1.0\linewidth]{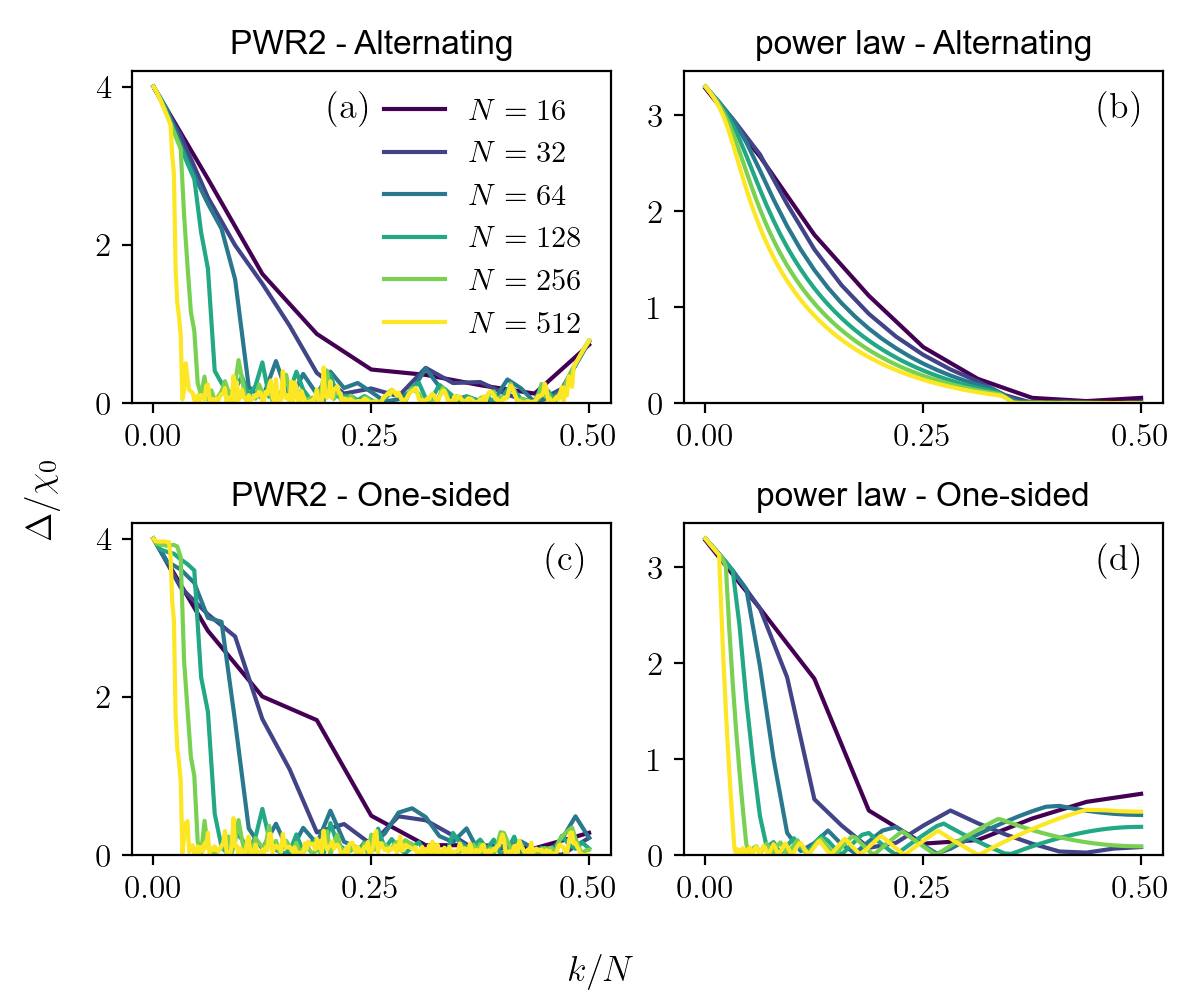}
    \caption{Spectral gap $\Delta/\chi_0$ between the collective subspace and the rest of Hilbert space, for the transformed Hamiltonian in Eq. (\ref{eq:transf_hami}), which is initial-state dependent. Gap is shown for various systems sizes as a function of the ratio of spin-flips in the initial state $k/N$. (a) and (c) correspond to PWR2 graph, while (b) and (d) to power-law decaying interactions with $\alpha=1$. Top row corresponds to alternating spin-flips, and bottom row to one-sided spin flips. In all cases, the gap decreases to zero for increasing $k/N$ (and does so faster as $N$ increases)  }
    \label{fig:spectral_gaps}
\end{figure}

With the generalized coupling graph construction in place, we can analyze the properties of the spectral gap for each Hamiltonian of interest (power-law and PWR2) and for different flipping patterns (alternating and one-sided). Results are shown in Fig. \ref{fig:spectral_gaps}. For the cases were we had identified a collective-to-scrambling crossover, we observe that the spectral gap decays as the number of spin flips $k$ increase. The decay is sharp and occurs at finite value of $k/N$, after which the gap remains close to 0, apart from certain revivals which are akin to those observed in Fig. \ref{Fig:Revivals_and_late_EE}. The results indicate that the spectral gap can only protect states with a small number of flips, beyond which the dynamics is expected to scramble beyond the symmetric subspace. The exception, as before, is the power-law alternating case, Fig. \ref{fig:spectral_gaps} (b), in which the gap decays smoothly and remains finite until $k/N\simeq 0.4$. We recall that this case was also where no significant early-time growth was found in the semiclassical sensitivity analysis. 
These numerical results agree with the Lyapunov analysis of Fig. \ref{fig:Lyapunov_exp} in determining the specific combination of coupling graph and flipping pattern leading to scrambling. 

In addition, we found that the spectral gap and Lyapunov exponent analyses show a `critical' ratio of $k/N$ for which the crossover takes place that gets smaller as system size increases. We analyze this phenomenon quantitatively by studying the scaling of such critical value of spin flips, which we call $k_\mathrm{\ast}$, as a function of $N$.
Remarkably, we find that both independent analysis yield the same result: the critical number of spin flips scales as $k_{\ast}\sim \sqrt{N}$. The results indicate that as system size increases, the collective phase is robust to a finite (but subextensive) number of spin flips. 
We stress that this analysis is carried out in systems of several hundreds of particles, giving us confidence on the scaling results. This efficient access to large system sizes would be unfeasible for direct simulation of quantum dynamics, but instead is enabled by the generalized one-axis-twisting construction of Eq. (\ref{eq:hami_goat}).

\begin{figure}
    \centering
    \includegraphics[width=1.0\linewidth]{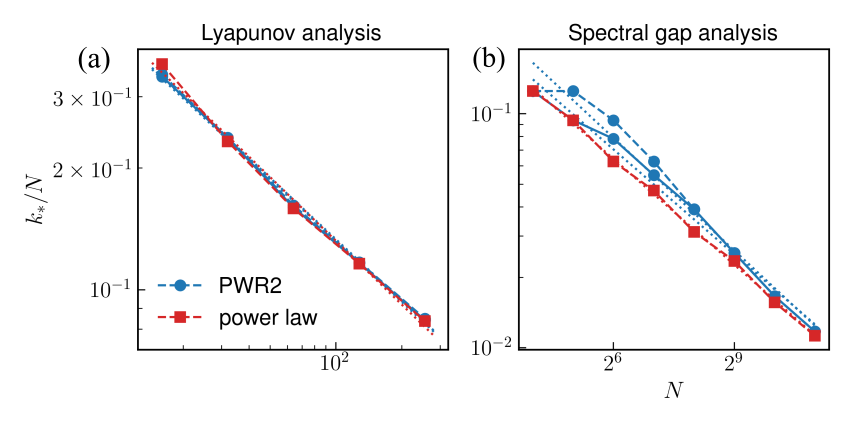}
    \caption{(a) Critical ratio $k_{\ast}/N$ as a function of system size $N$, obtained from the Lyapunov analysis. We define $k_{\ast}/N$ as the ratio at which $\lambda_L$ in \figref{fig:Lyapunov_exp} exceeds $\sim 3$. Results are shown for the PWR2 and PL graphs, for alternating (solid) and one sided (dashed) spin flip configurations. The alternating-spin-flip results for the PL graph are omitted because no corresponding increase in the Lyapunov exponent is observed. The dotted lines show fits of the form $\frac{k_{\ast}}{N} \propto N^{-\beta}$. For the
    PWR2 graph, the fitted exponents $\beta = 0.50 \pm0.01$ for both one-sided and alternating spin flips. For the PL graph with one-sided spin flips, $\beta = 0.52 \pm 0.02$. These results are consistent with $\frac{k_{\ast}}{N} \propto N^{-0.5}$ or equivalently $k_{\ast}\sim \sqrt{N}$.
    (b). Critical ratio $k_{\ast}/N$ as a function of system size $N$ obtaied from the spectral gap analysis. The symbols and line styles follow the same conventions as in panel (a), and the dotted lines show the corresponding fits. For the PWR2 graph, the fitted exponents are $\beta = 0.53 \pm 0.03$ and $0.50 \pm 0.02$ for one-sided and alternating spin flips respectively. For the PL graph with one-sided spin flips, $\beta = 0.50 \pm 0.01$ 
    }
    \label{fig:critical_spin_flips}
\end{figure}
\section{Conclusion and outlook} \label{sec:conclusion}
In this work, we presented a class of long-range interacting spin models that host a crossover between collective to scrambled dynamics as the initial state is changed. The models studied are U(1) symmetric XY chains, with no free parameters (e.g. no external fields), and the crossover is driven by the number of permutation-symmetry-breaking spin flips in the initial state is increased. While different flipping patterns lead to different dynamics in general, we find specific protocols that make the crossover more prominent, and in general observe that randomness smoothens out the emergence of scrambling. 

In particular, for the PWR2 and power-law decay coupling graphs, we showed that one-sided and alternating spin-flip patterns generate two distinct dynamical regimes. Among these, the alternating pattern with $N/2 $ spin flips is especially significant because it is an exact eigenvector of the circulant adjacency matrix of the graph, allowing the growth of classical sensitivity to be understood cleanly through the linearized equations of motion. This provides a simple geometry-based criterion for diagnosing chaotic dynamics in the XY Hamiltonian: one can infer the relevant classical sensitivity from the spectral properties of the flipped initial state. This perspective naturally extends our earlier work on phases and criticality in sparse long-range systems \cite{gunning2025geometrydriventransitionssparselongrange}.

Our results are supported by a range of diagnostics and analysis spanning quantum and classical dynamics and kinematics. These tools allow us to connect the emergence of the crossover to the graph-theoretic properties of the coupling graph, as well to access large system sizes, otherwise inaccessible to direct numerical simulation. 

The framework developed here naturally suggests several directions for future work. First, while we focused on the PWR2 and power-law coupling geometries, the spectral and semiclassical tools introduced in this work provide a route to systematically analyse and optimise interaction graphs for information scrambling. In particular, it should be possible to formulate graph-design principles that maximise Lyapunov growth, enlarge the scrambling regime, or engineer desired crossovers between collective and chaotic dynamics. These ideas could be extended beyond one-dimensional translationally-invariant systems to more general sparse long-range networks and higher-dimensional lattices.

A second important direction concerns experimental realisation. The models studied here are directly relevant to a variety of quantum simulation platforms with programmable long-range interactions, including neutral-atom arrays, trapped ions, and cavity-mediated spin systems. Importantly, the crossover does not rely on tuning Hamiltonian parameters: the interaction graph is fixed, and the transition is driven entirely by simple local spin-flip operations applied to the initial state. Combined with the absence of free parameters and the accessibility of probes such as collective correlation functions, this makes the phenomenon particularly attractive for near-term experiments. As the collective-to-scrambling crossover may be viewed as a transition in classical simulability, it also provides a experimentally accessible benchmark for analog quantum simulators in which the underlying Hamiltonian remains unchanged while the complexity of the dynamics is controlled through state preparation alone.

Finally, our spectral analysis suggests a broader connection between scrambling dynamics and transport on signed graphs. For collective initial states, the Laplacian gap is directly related to the mixing time of an associated random walk. The spin-flip transformation introduced here generates signed interaction graphs whose Laplacians continue to govern the separation between collective and non-collective sectors. Understanding the resulting dynamics in terms of random walkers on signed networks may therefore provide a unified framework linking information scrambling, graph topology, and spectral transport phenomena. Developing this connection more formally could yield new analytical tools for predicting many-body dynamics directly from graph-theoretic properties





\begin{acknowledgments}
 We thank Gregory Bentsen,  Monika Schleier-Smith, Zlatko Papi\'c, Ryan Connor and Teddy Ong for helpful discussions. This work was supported by the EPSRC through the QQQS programme grant (EP/Y01510X/1) and by the QCi3 hub (EP/Z53318X/1).
 PMP acknowledges support from the Royal Society through a University Research Fellowship (Grant No. URF\textbackslash R1\textbackslash252030). In compliance with EPSRC's open access initiative, the data in this paper will be available from \cite{data_open_access}.
 \end{acknowledgments}

\bibliography{References.bib}
\pagebreak
\widetext
 
 \newpage 
\begin{center}
\vskip0.5cm
{\Large Appendices}
\end{center}
\vskip0.4cm

\setcounter{section}{0}
\setcounter{equation}{0}
\setcounter{figure}{0}
\setcounter{table}{0}
\setcounter{page}{1}
\renewcommand{\theequation}{S\arabic{equation}}
\renewcommand{\thefigure}{A\arabic{figure}}
\renewcommand{\thesection}{A\arabic{section}}
\renewcommand{\thesection}{A\arabic{section}}
\section{QFI and Entanglement Dynamics Averaged over Spin-Flip Configurations}
\label{App:Avg_QFI_EE}
\begin{figure}[h!]
    \centering
    \includegraphics[width=0.70\linewidth]{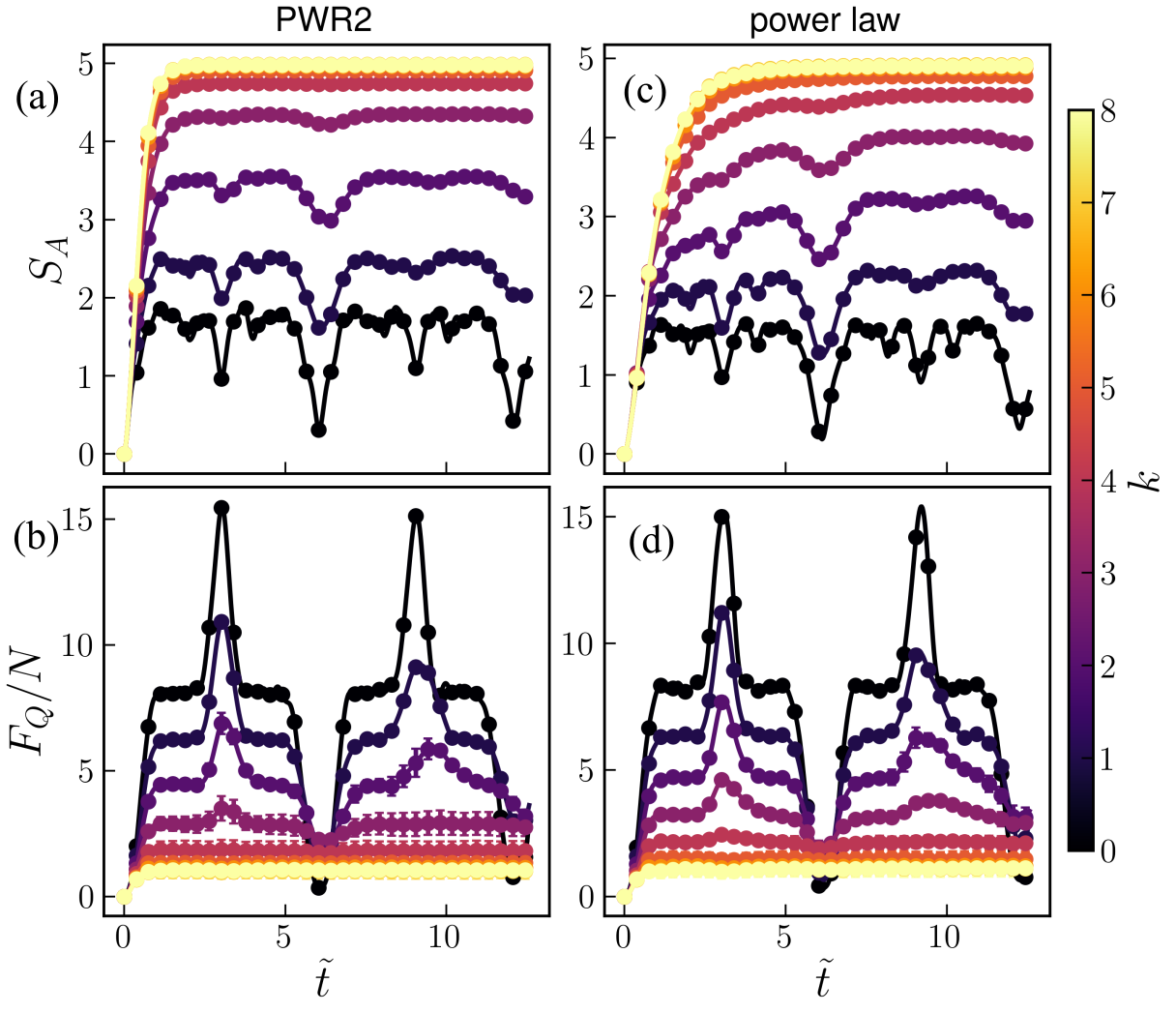}
    \caption{Evolution of bipartite entanglement entropy $S_A$ and quantum Fisher information density $F_Q/N$ under the $XY$ Hamiltonian given in \eqeqref{eq:XY Hamiltonian} for $N=16$. Panels (a,b) correspond to the PWR2 graph, while panels (c,d) correspond to the PL graph with $\alpha=1.0$. The initial state is $x$-polarised, with $k$ spins flipped in the $x$ basis, as indicated by the colour bar.  (a) PWR2 periodic graph and (c) power law decaying graph with $\alpha=1.0$ for $N=16$ and different numbers of spin flips $k$. For each $k$, we average over all $\binom{N}{k}$ distinct configurations when $\binom{N}{k}<200$, and otherwise, over 200 uniformly samples distinct configurations. We take $|A|=N/2$, and time is normalised according to \eqeqref{eq:time_norm}. Error bars indicate the variation across configurations and are smaller than the markers when not visible.}
    \label{fig:avg_QFI_EE}
\end{figure}

The dynamics of the quantum Fisher information (QFI) and entanglement entropy (EE) can be studied using initial states with different spin-flip patterns. For a system of (N) spins with (k) flipped spins, there are $\binom{N}{k}$ distinct spin-flip configurations. To determine whether different configurations with the same value of (k) exhibit similar dynamics, we average the QFI and EE over all $\binom{N}{k}$ configurations when $\binom{N}{k}<200$. Otherwise, we sample 200 distinct configurations uniformly at random. The resulting dynamics are shown in \figref{fig:avg_QFI_EE} for the PWR2 model in panels (a) and (b), and for the PL coupling graph in panels (c) and (d). The error bars indicate the variation across the sampled spin-flip configurations. Their increase for larger values of (k) is consistent with the configuration-dependent behaviour discussed in the main text.
\section{Late time studies of entanglement entropy for one sided spin flips}
\label{App:late_time_OS}
\begin{figure}
    \centering
    \includegraphics[width=0.7\linewidth]{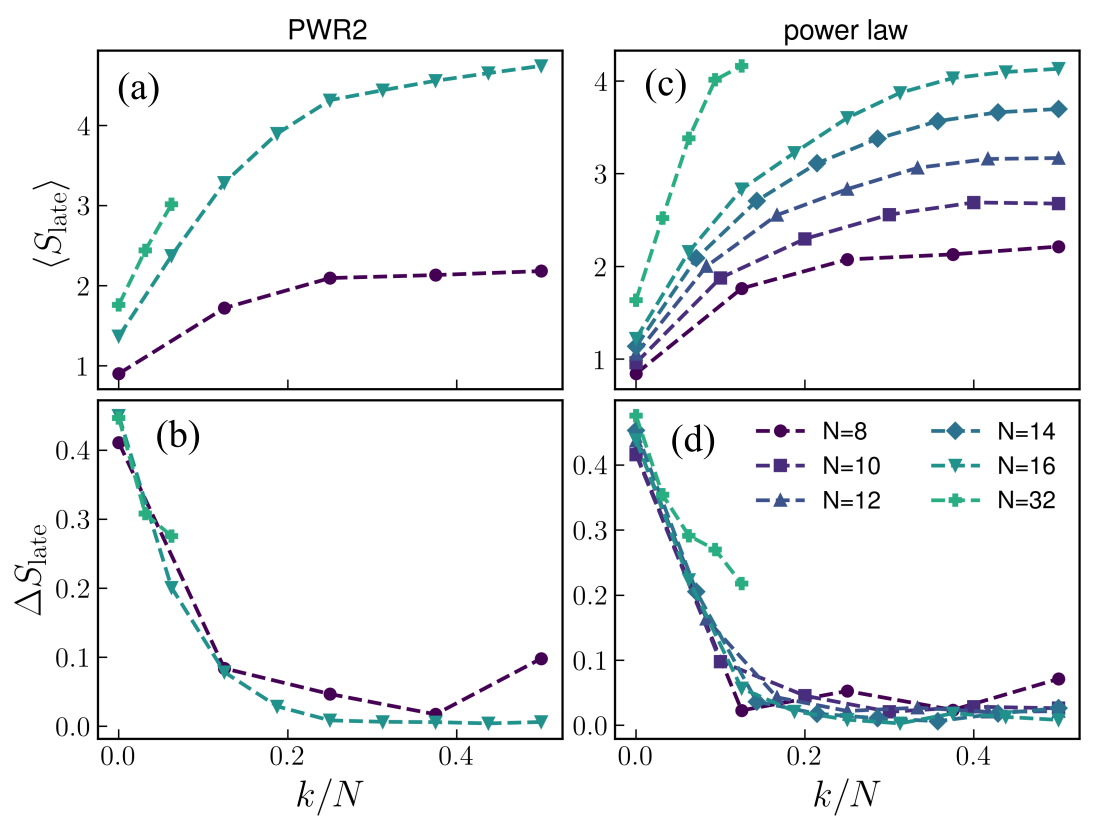}
    \caption{Average late-time entanglement entropy, $\langle S_{\mathrm{late}}\rangle$, and its temporal standard deviation, defined in \eqeqref{eq:standard deviation of S}, under dynamics generated by the $XY$ Hamiltonian in \eqeqref{eq:XY Hamiltonian}. Late times are defined by $\tilde{t}>10$, and the results are plotted as functions of the fraction of spins flipped in the $x$ basis, $k/N$. Panels (a) and (b) show results for the PWR2 graph, whereas panels (c) and (d) show results for the PL coupling graph with $\alpha=1.0$. For the PWR2 graph, we consider $N=8,16,$ and $32$; for the PL graph, we consider $N=8,10,12,14,16,$ and $32$. For $N=32$, fewer data points are included because the rapid growth of entanglement limits the accessible simulation times in classical tensor-network calculations.}
    \label{fig:Revivals_OS}
\end{figure}

In addition to the results presented in the main text, we calculate the late-time entanglement entropy and its temporal fluctuations, as defined in \eqeqref{eq:standard deviation of S}, for initial states generated using one-sided spin flips. The results are shown in \figref{fig:Revivals_OS}. Panels (a) and (b) correspond to the PWR2 graph, while panels (c) and (d) correspond to the PL graph. 

As expected, the average late-time entanglement entropy, $\langle S_{\mathrm{late}}\rangle$, increases with fraction of flipped spins $k/N$, In contrast to the alternating-spin-flip protocol, no decrease is observed at $k=N/2$. 
Similarly, the temporal fluctuations of the entanglement entropy decrease as $k/N$ increases, with no rise at $k=N/2$. 
\section{Semi-Classical Equations of Motion} \label{app:classical_eoms}
The Heisenberg Hamiltonian is written as
\begin{align}
    H = \frac{1}{2S} \sum_{i,j} \chi_{i,j}(S_i^{x}S_{j}^{x}+ S_i^{y}S_{j}^{y}),
\end{align}
where each site is considered to be a spin $S$ particle. According to Heisenberg's equations of motion,
\begin{align}
    \frac{dA(t)}{dt} &= \frac{i}{\hbar} [H,A] + \Big(\frac{\partial A}{\partial t}\Big), \\
    \frac{dA(t)}{dt} &= \frac{i}{\hbar} [H,A] 
    \label{eq:HEOM}
\end{align}
where $A$ is an operator. Thus for each site $k$ one may write
\begin{align}
    \frac{d S_k^{x,y,z}}{dt} = \frac{i}{2S}\Big[\sum_{i,j}\chi_{i,j}(S_i^{x}S_{j}^{x}+ S_i^{y}S_{j}^{y}),S_{k}^{x,y,z}\Big].
\end{align}
We then solve it separately for each component. Starting with the $x$ direction, we get
\begin{align}
    \frac{d \langle S_k^{x} \rangle}{dt} = \frac{1}{2S}\sum_{i} \chi_{i,k}(\langle S_{i}^{y}S_{k}^{z}\rangle + \langle S_{k}^{z}S_{i}^{y}\rangle).
\end{align}
We consider the mean-field ansatz $\langle AB \rangle = \langle A \rangle \langle B \rangle $, and this gives us
\begin{align}
    \frac{d \langle S_k^{x} \rangle}{dt} = \frac{1}{2S}\sum_{i} \chi_{i,k}(\langle S_{i}^{y}\rangle \langle S_{k}^{z}\rangle + \langle S_{k}^{z} \rangle \langle S_{i}^{y}\rangle).
\end{align}
We define $\vec{r_k} = (X_k, Y_k, Z_k) = \frac{\vec{S_k}}{S}$, yielding us
\begin{align}
    \dot{X}_k &= \frac{1}{2}\sum_{i} \chi_{i,k}(Y_i Z_k + Z_k Y_i)\\
    \dot{X}_k & = \sum_{i} \chi_{i,k}(Y_i Z_k).
\end{align}
Similarly, we find that 
\begin{align}
\dot{Y}_k = -\sum_{i} \chi_{i,k}(X_iZ_k).
\end{align}
These calculations are similar to that for the all to all coupling graph. However, the $z$ component in the all to all coupling graph goes to zero, which is not the case in any generic graph. Following the same steps, we calculate the $Z$ component and get the following:
\begin{align}
    \dot{Z}_{k} = \sum_{i}\chi_{i,k} (X_iY_k - Y_iX_k).
\end{align}
Thus we have the set of $3N$ different equations which can be solved numerically
\begin{align}
\label{3n_eq_supp}
   \dot{X}_k & = \sum_{i} \chi_{i,k}(Y_i Z_k), \nonumber \\
   \dot{Y}_k &= -\sum_{i} \chi_{i,k}(X_iZ_k), \nonumber \\
   \dot{Z}_{k} &= \sum_{i}\chi_{i,k} (X_iY_k - Y_iX_k).
\end{align}
\section{Calculation of Lyapunov exponent} \label{app:lyapunov_calc}
Once we have the $3N$ differential equations, they can be solved numerically to evaluate the semiclassical OTOC defined in the main text \eqeqref{eq:OTOC}. For a given initial configuration with $k$ spin flips, we introduce a small perturbation at site $i$ and monitor its effect on site $j$ as a function of time. The OTOC therefore depends on the graph distance between the two sites $r_{ij}$. 
\subsection{Graph Distance}
The graph distance $r_{ij}$ is defined as the shortest path between vertices $i$ and $j$. For unweighted graphs, $r_{ij}=1$ if there is an edge connecting $i$ and $j$; more generally, $r_{ij}=D$, where $D$ is the minimum number of edges that must be traversed to reach $j$ starting from $i$. 
For the PWR2 graph with periodic boundary conditions, the maximum graph distance is $r_{max} = \lceil \frac{1}{2} \log_2{N} \rceil$. 
For the power law decaying graph with $\alpha = 1.0$ and periodic boundary conditions, we instead consider a weighted graph distance. Each edge is assigned a weight $w_{ij} = |i-j|^{\alpha}$, and the distance  $r_{ij}$ is defined as the minimum total weight along all possible paths connecting $i$ and $j$. 
For example, for $N=8$ and $\alpha=1$, the distance between sites  $i=0$ and $j=4$ is $r_{ij}=4$. The direct connection has weight $4$, while the sequential path $0 \to 1 \to 2 \to 3 \to 4$ has total weight $1+1+1+1=4$; hence the shortest-path distance is $4$.
\subsection{Fitting the curves}
\begin{figure}
    \centering
    \includegraphics[width=0.7\linewidth]{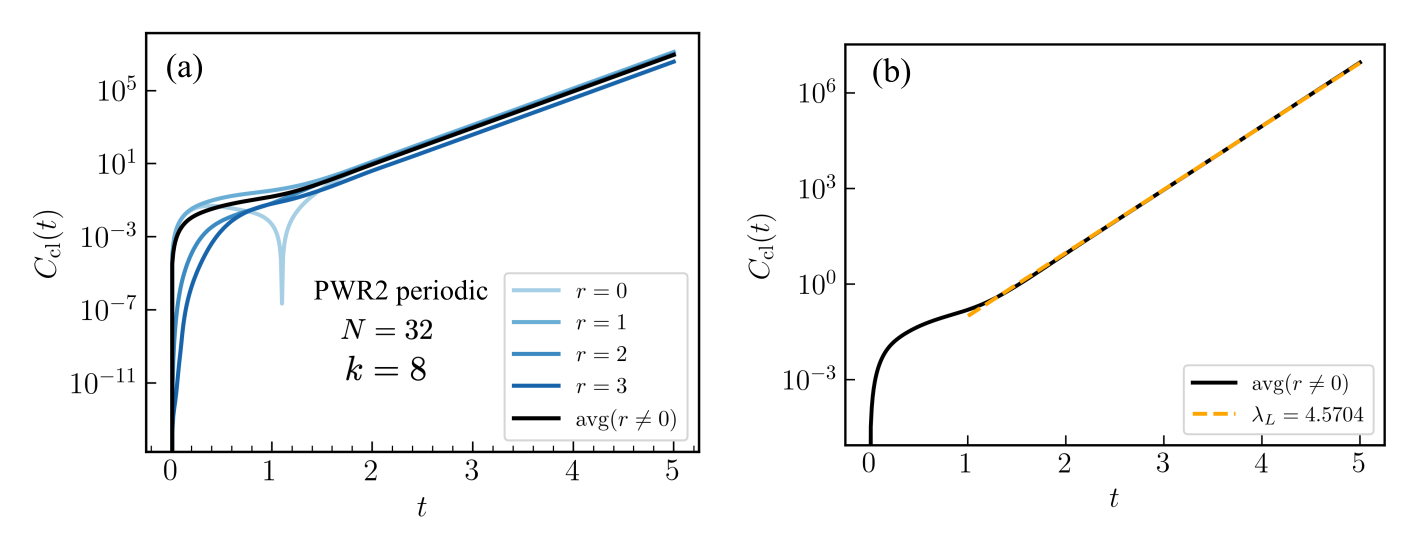}
    \caption{(a) The semi-classical OTOC $C_{cl}(t)$ \eqeqref{eq:OTOC} is plotted as a function of time for different graph distances $r$, for $N=32$ with $k=8$ spin flips. Different values of $r$ are shown in fading blue colours. The OTOC averaged over all non-zero graph distances is shown as the black solid line. (b). Exponential fit of the averaged curve in the time window $1 \lesssim \chi_0 t \lesssim 5$, shown as the orange dashed line. The Lyapunov exponent $\lambda_L$ is extracted from its slope. }
    \label{fig:lyp_extraction}
\end{figure}

After numerically simulating the OTOCs for different graph distances $r_{ij}$, we average over all spin pairs separated by the same graph distance. This yields a set of curves for $C_{\mathrm{cl}}(t)$ as a function of time for each $r_{ij} \neq 0$ (we exclude $r=0$, which corresponds to the same site). An example for $N=32$ and $k=8$ spin flips on the PWR2 graph is shown in \figref{fig:lyp_extraction}(a). 
We then average over all graph distances to obtain a single representative curve. The resulting curve is shown as the black line in \figref{fig:lyp_extraction}(a). 
For short times, the OTOC exhibits polynomial growth, $C_{\mathrm{cl}}(t) \propto (\chi_0 t)^{2 r_{ij}}$ as mentioned in the main text, valid in the regime $\chi_0 t \lesssim 1$. To extract the Lyapunov exponent, we instead focus on the intermediate-time window $1 \lesssim \chi_0 t \lesssim 5$, where exponential growth is expected.
Following Ref.~\cite{Greg_tree_2019}, we assume
\begin{equation}
    C_{\mathrm{cl}}(t) = \frac{e^{\lambda_L t}}{N^a}.
\end{equation}
Taking the logarithm gives
\begin{align}
    \log C_{\mathrm{cl}}(t) 
    &= \log\!\left(e^{\lambda_L t}\right) - \log\!\left(N^a\right) \\
    &= \lambda_L t - a \log N.
\end{align}
Thus, $\lambda_L$ can be extracted from the slope of $\log C_{\mathrm{cl}}(t)$ in the exponential growth regime. \figref{fig:lyp_extraction}(b) shows a representative fit (orange dashed line), from which $\lambda_L$ is obtained. 
\section{Scaling of Lyapunov exponents}
\label{app:Scaling_lyp}
\begin{figure}
    \centering
    \includegraphics[width=0.7\linewidth]{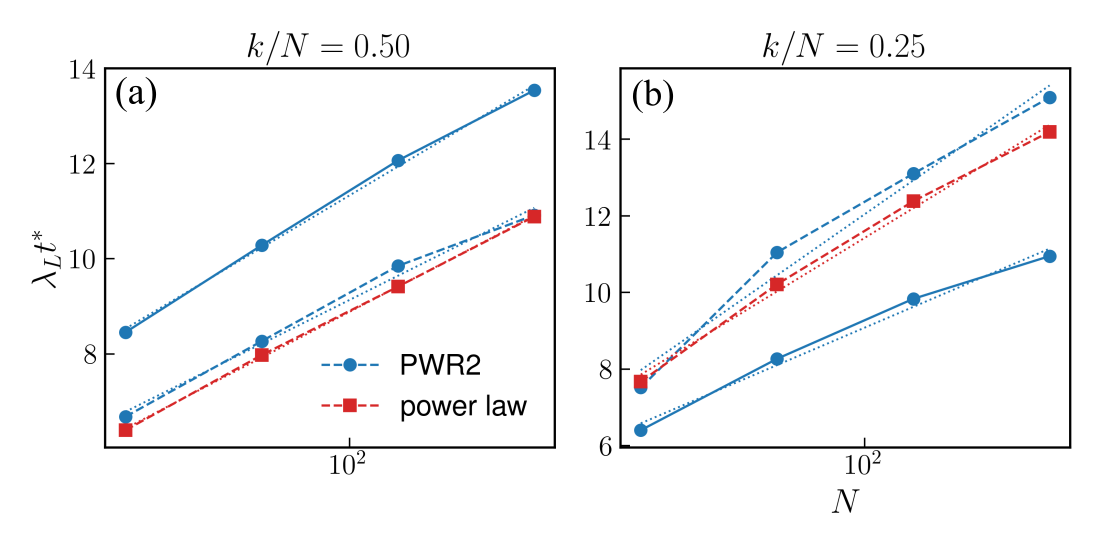}
    \caption{Scaling of the scrambling time with system size for (a) $\frac{k}{N}=0.5$ and (b) $\frac{k}{N}=0.25$ . The quantity $\lambda_L t^{*}$ is plotted as a function of the system size $N$ for $k=\frac{N}{2}$ spin flips. Results are shown for the PWR2 graph and for the power-law decaying, each with both alternative and one sided spin flip protocols. For the PWR2 graph initialised in the Neel state in the x basis, with the alternative flipping protocol, we observe $\lambda_L t^{*} \propto \log(N)$, as confirmed by the best fit line (dotted line). The same logarithmic scaling of $\lambda_L t^{*}$ with $N$ is also observed for the one-sided spin flips (dashed line).
    For the power law decaying graph, the logarithmic scaling of $\lambda_L t^{*}$ with $N$ is observed for the one-sided spin flips. However, the scrambling behaviour differs between two spin flipping protocols. In the alternative spin-flip case, the Lyapunov exponent $\lambda_L$ does not increase with system size (see \figref{fig:Lyapunov_exp}(c)) and the classical sensitivity $C_{cl}$ fails to reach $O(1)$, indicating the absence of fast scrambling in that regime.}
    \label{fig:scaling_lyp}
\end{figure}

For the maximal number of spin flips, $k=N/2$, we compute $t^{*}$ at which $C_{cl}(t^{*})=1$. We then examine the scaling of $\Lambda_Lt^{*}$ with system size $N$, as shown in \figref{fig:scaling_lyp}(a). Our semiclassical analysis predicts $t^{*} \sim \log(N)$ for both power law decay and PWR2 graphs. However, for the power-law decaying graph we do not observe any growth of the Lyapunov exponent (see \figref{fig:Lyapunov_exp} (c)), and $C_{cl}$ fails to reach $O(1)$ for alternating spin flips. The absence of exponential growth indicates that the power-law decaying graph does not exhibit fast scrambling dynamics in this regime. 
We also calculate the scaling of $\lambda_Lt^{*}$ with system size $N$ for $k=N/4$, shown in \figref{fig:scaling_lyp}(b). In this case, the scaling is less pronounced than for $k/N = 0.5$. Nevertheless, it suggests that increasing the fraction of spin flips enhances scrambling behaviour. 
\section{Derivation of the linearised Equations of Motion} \label{app:linear_eoms}
The $3N$ coupled equations of motion given in \eqref{3n_eq_supp} admit a fixed point of the form
$(X_l, Y_l, Z_k) = (\sigma_l, 0, 0), \quad \sigma_l \in \{\pm 1\}$,
for all sites $l$. Classical sensitivity is then characterized by the growth of small perturbations around this fixed background.
We linearize the dynamics by writing
\begin{align}
    X_l &= \sigma_l + \delta x_l, \nonumber\\
    Y_l &= \delta y_l, \nonumber \\
    Z_l &= \delta z_l,
\end{align}
and retain only terms linear in the perturbations. This yields the linearized equations of motion
\begin{align}
\label{linearised_eqns}
    \dot{\delta x}_l &= 0, \nonumber\\
    \dot{\delta y}_l &= -m_l \, \delta z_l, \nonumber \\
    \dot{\delta z}_l &= m_l \, \delta y_l - \sigma_l \sum_{i}\chi_{li} \, \delta y_i,
\end{align}
where 
$m_l = (\chi \sigma)_l = \sum_{i}\chi_{li} \sigma_i$.
Since the underlying graph is periodic, the coupling matrix $\chi$ is circulant, a property that will allow us to diagonalize the linearized dynamics in Fourier space.

The alternating background, defined by $\sigma_l = (-1)^l$, is an eigenvector of the circulant matrix $\chi$, with eigenvalue $\lambda_{\pi} \equiv \lambda_{q=N/2}$. This implies
$m_l = (\chi \sigma)_l = \lambda_{N/2} \sigma_l$. 
Substituting into \eqref{linearised_eqns}, we obtain
\begin{align}
\label{linearised_eqns_withlambda}
    \dot{\delta x}_l &= 0, \nonumber\\
    \dot{\delta y}_l &= -\lambda_{N/2} \sigma_l \, \delta z_l, \nonumber \\
    \dot{\delta z}_l &= \sigma_l\left(\lambda_{N/2}\delta y_l - (\chi \, \delta y)_l \right).
\end{align}

Taking a time derivative of $\delta y_l$, we find
\begin{align}
\label{eq:doubderivative}
    \delta \ddot{y}_l &= -\lambda_{N/2} \sigma_l \, \delta \dot{z}_l \nonumber \\
    &= \lambda_{N/2} (\chi \, \delta y)_l - \lambda_{N/2}^2 \, \delta y_l.
\end{align}
We now introduce the Fourier decomposition
\begin{align}
    \delta y_l = \frac{1}{\sqrt{N}} \sum_q y_q \, e^{\frac{2\pi i ql}{N}},
\end{align}
where $q$ labels the Fourier modes. Since $\chi$ is circulant, its eigenvectors are precisely these Fourier modes, with eigenvalues $\lambda_q$. In particular, the alternating configuration corresponds to the mode $q = N/2$, with eigenvalue $\lambda_{\pi}$.
Substituting into \eqref{eq:doubderivative}, each Fourier mode decouples and evolves independently as
\begin{align}
\label{eq: Supp_LE}
    \ddot{y}_q = \lambda_{N/2}(\lambda_q - \lambda_{N/2}) \, y_q.
\end{align}
\section{Eigenvectors and Fourier modes of Circulant Matrices}
\label{app:eigenvectors_Uniform_OS}
\subsection{Uniform Background}

\begin{figure}
    \centering
    \includegraphics[width=0.75\linewidth]{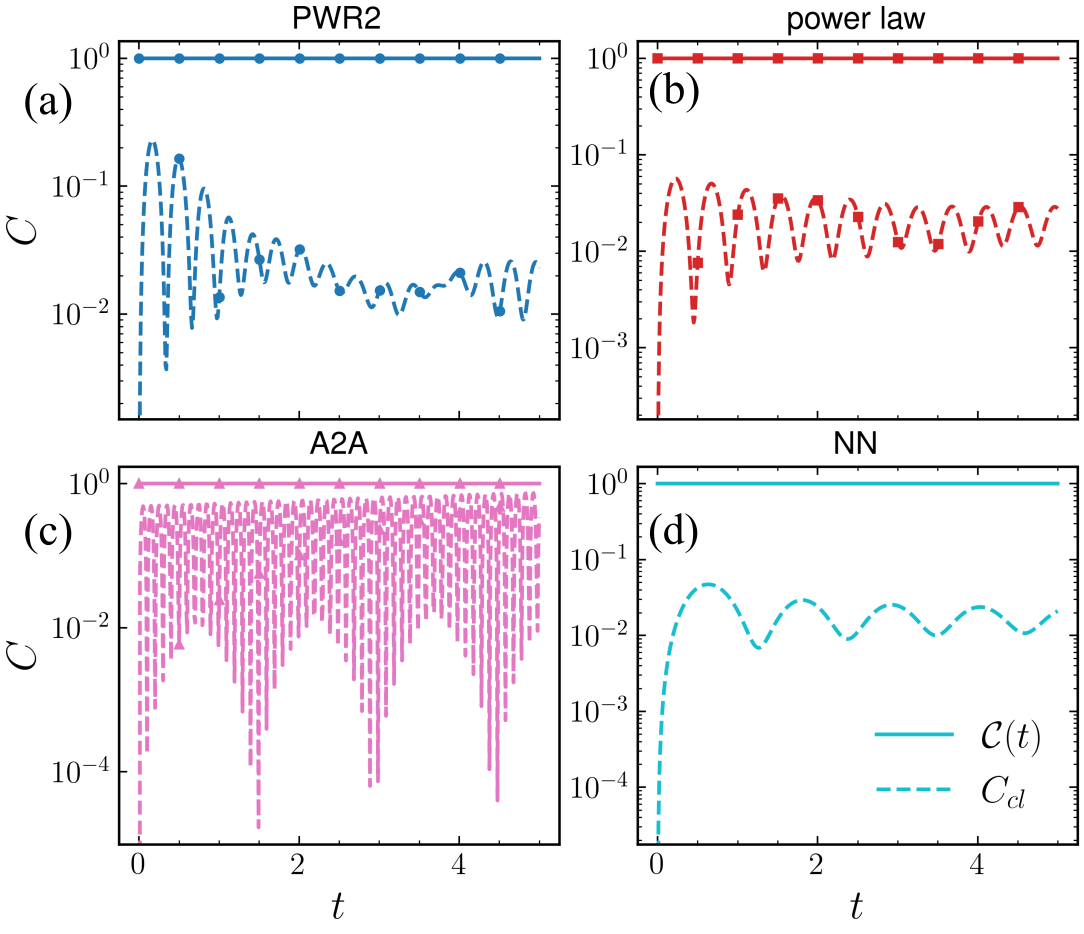}
    \caption{Classical sensitivity $C_{\mathrm{cl}}$ (dashed), defined in \eqeqref{eq:OTOC}, and $\mathcal{C}(t)$ (solid), defined in \eqeqref{eq:recipe_otoc}, as functions of time $t$ for different periodic coupling matrices: (a) PWR2, (b) power-law decay with $\alpha = 1.0$, (c) all-to-all (A2A), and (d) nearest-neighbor graphs. As expected, there is no increase in the classical sensitivity.}
    \label{fig:Recipe_OTOC_uniform}
\end{figure}

In the main text, we showed how the classical sensitivity derived from the linearised equations of motion depends on the alternating and one sided spin-flip background. Here, we analyse the uniform background case $\psi_{\mathrm{BG}} = (+1,+1,\cdots,+1)$. As mentioned in the main text, this is an eigenvector of the circulant matrices we consider and corresponds to the $q=0$ mode. Hence, the linearised equation of motion in in \eqeqref{eq: Supp_LE} can be written as
\begin{align}
\label{eq: Supp_LE_mode0}
    \ddot{y}_q = \lambda_{0}(\lambda_q - \lambda_{0}) \, y_q,
\end{align}
and the total sensitivity can be defined as $\mathcal{C}(t) \sim \sum_{q} |c_q|^2e^{2 \Gamma_qt}$, with $\Gamma_{q} = \sqrt{\lambda_0(\lambda_q - \lambda_0)}$.
The results are shown in \figref{fig:Recipe_OTOC_uniform} for $N=32$ and for all coupling geometries. As expected, we observe no growth in $\mathcal{C}(t)$. 
\subsection{One Sided Spin Flip Background}
\begin{figure}
    \centering
    \includegraphics[width=0.5\linewidth]{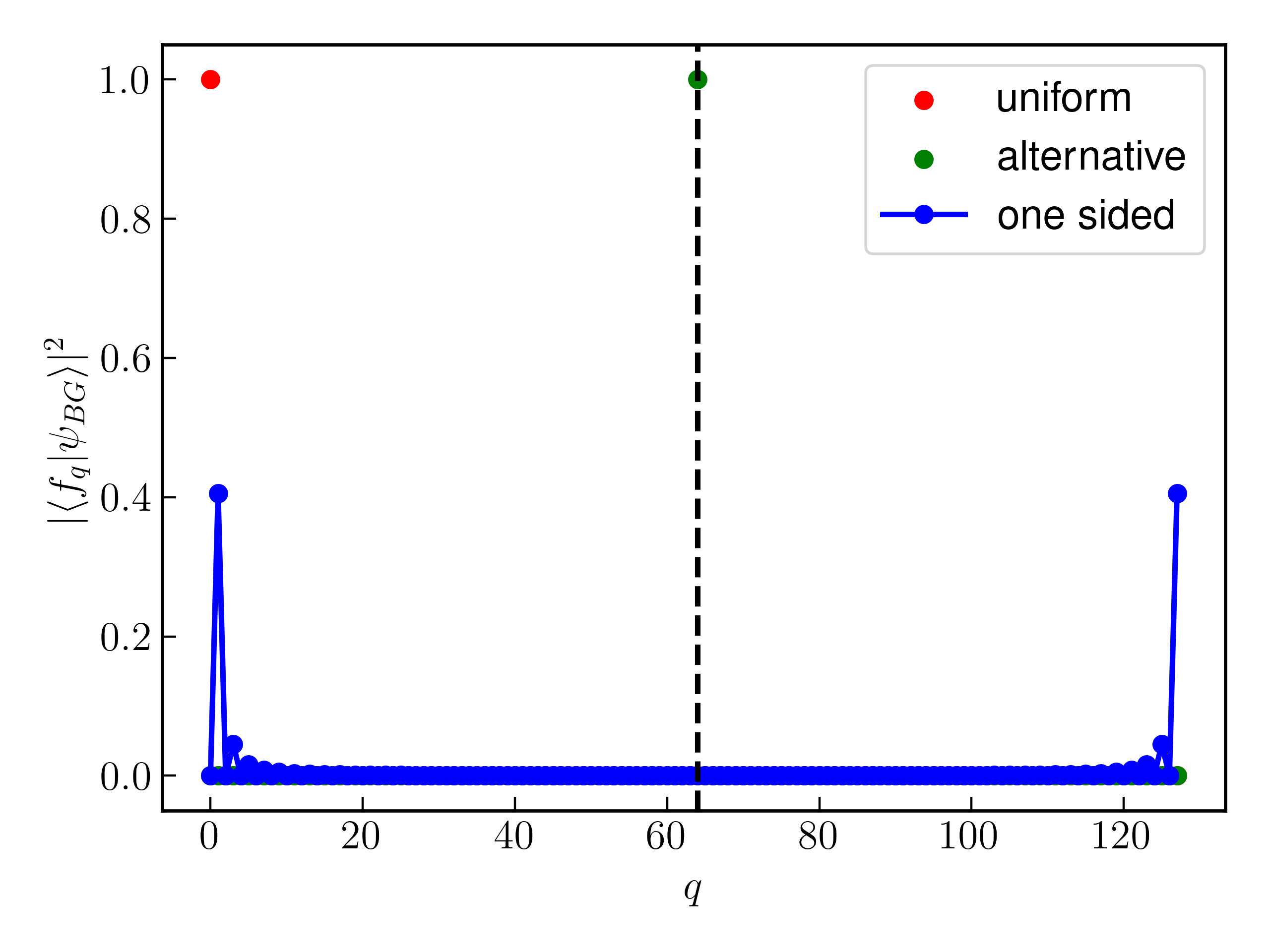}
    \caption{Overlap of the background spin configuration $\psi_{\mathrm{BG}}$ with the eigenvectors $f_q$ of the circulant matrix in \eqeqref{eq:eigenvector_cm}, shown as a function of mode index $q$. The red point corresponds to the uniform background $\psi_{\mathrm{BG}} = (+1,+1,\dots,+1)$, which overlaps perfectly with the $q=0$ mode. The green point corresponds to the alternating background $\psi_{\mathrm{BG}} = (+1,-1,+1,-1,\dots)$, which overlaps perfectly with the $q=N/2$ mode. The blue curve shows the one-sided flip background $\psi_{\mathrm{BG}} = (-1,-1,\dots,+1,+1,\dots)$, whose largest overlaps occur at $q=1$ and $q=N-1$. Here $N=128$. The vertical dashed line marks $q=N/2$ as a guide to the eye.}
    \label{fig:overlaps}
\end{figure}

As explained in the main text, the uniform background and the alternating spin background are eigenvectors of the circulant matrix. By contrast, the one-sided spin-flip background is not an eigenvector. For a system of $N$ spins with $N$ even, it is defined as
\begin{align}
 \sigma_l = 
    \begin{cases}
       -1, & \mathrm{for}\, 0 \leq l < N/2 \\
+1,  & \mathrm{for}\, N/2 \leq l < N.
    \end{cases}
\end{align}
Its Fourier coefficients are
\begin{align}
    \sigma_q = \frac{1}{\sqrt{N}}\sum_{k=0}^{N-1} \sigma_l e^{\frac{-2 \pi i l}{N}},
\end{align}
which gives 
\begin{align}
\label{eq:Fourier_odd_spins}
 \sigma_q = 
    \begin{cases}
       0, & \mathrm{if}\,q \ \mathrm{is} \ \mathrm{even} \\
\frac{2}{\sqrt{N}}\frac{e^{\frac{i\pi q}{N}}}{i \sin{(\frac{\pi q}{N}})},  & \mathrm{for}\, N/2 \leq l < N.
    \end{cases}
\end{align}
\eqeqref{eq:Fourier_odd_spins} shows that the one-sided flip state has overlap only with odd Fourier modes. Among these, the dominant contributions come from $q=1$ and $q=N-1$. To illustrate this, we plot the overlap of the one-sided flip background $ \psi_{BG} = (-1,-1,\dots,+1,+1,\dots)$ with the eigenvectors of the circulant matrix in \eqeqref{eq:eigenvector_cm}. Figure~\ref{fig:overlaps} shows the result for $N=128$. The dominant modes are indeed $q=1$ and $q=N-1$ (blue). For comparison, we also show the uniform background and the alternating spin background, which overlap perfectly with $q=0$ (red) and $q=N/2$ (green), respectively.

\section{Signed graphs from spin flipping}
\label{app:spin_flips_gap}

We are interested in dynamics generated by the XY Hamiltonian on a given graph $\chi_{ij}$,
\begin{equation}
    H = \sum\limits_{ij}\chi_{ij} (X_i X_j + Y_i Y_j)
\end{equation}
and starting from an initial state which can be thought of as a fully polarized state along $+x$ and then acted on with a sequence of flips,
\begin{equation}
    \Ket{\Psi_k^{\balpha}(0)} = \mathcal{Z}_k^{\balpha} \Ket{+}^{\otimes N}\ \mathrm{where}\ \mathcal{Z}_k^{\balpha} = Z_{\alpha_1}Z_{\alpha_2}\ldots Z_{\alpha_k}
\end{equation}
where $k$ is the number of flips and $\balpha$ is a vector containing the indices of the spins that have been flipped. Time evolution starting from this state and following $H$ leads to

\begin{equation}
    \Ket{\Psi_k^{\balpha}(t)}=e^{-i H t}\mathcal{Z}_k^{\balpha} \Ket{+}^{\otimes N} = \mathcal{Z}_k^{\balpha} e^{-i H_k^{\balpha} t}\Ket{+}^{\otimes N}
\end{equation}
where 
\begin{equation}
    H_k^{\balpha} = \mathcal{Z}_k^{\balpha} H \mathcal{Z}_k^{\balpha}. 
\end{equation}

Since $\mathcal{Z}_k^{\balpha}$ is local, it follows that it cannot change the entanglement properties of the state. So, it makes sense to focus on the dynamics generated by the transformed Hamiltonian $H_k^{\balpha}$. Furthermore, note that 
\begin{equation}
    Z_l(X_i X_j  + Y_i Y_j)Z_l = \left\{ \begin{array}{cc}
        -(X_i X_j  + Y_i Y_j) & (l=i\ \mathrm{or}\ l=j)  \\
         X_i X_j  + Y_i Y_j & (l\neq i,\: l\neq j) 
    \end{array}\right.
\end{equation}
This means that the action of $Z_l$ changes the interaction graph as
\begin{equation}
    \chi_{ij} \rightarrow (-1)^{\delta_{il} + \delta_{jl}}\chi_{ij}.
\end{equation}

Focusing on the $N\times N$ adjacency matrix $A_{ij}=\chi_{ij}$, then what happens is that 
\begin{equation}
    A\rightarrow D_l A D_l
\end{equation}
with $(D_l)_{ij}=(-1)^{\delta_{il}}\delta_{ij}=\mathrm{diag}(1,1,...,-1,..,1,1)$. The action of all $Z_l$'s commute with each other, so in the end we have that $H_k^{\balpha}$ has a modified adjacency matrix $A_k^{\alpha}$ of the form
\begin{equation}
    A_k^{\alpha} = \mathcal{D}_k^{\balpha}A\mathcal{D}_k^{\balpha},\ \mathrm{where}\ \mathcal{D}_k^{\balpha}=D_{\alpha_1}D_{\alpha_2}\ldots D_{\alpha_k}.
    \label{eq:A_k_alpha}
\end{equation}
\end{document}